\documentclass[longauth]{aa}  

\usepackage{txfonts}

\usepackage{booktabs}
\usepackage{subfigure}
\usepackage{graphicx}
\usepackage{amssymb}
\usepackage{textcomp}
\usepackage{url}
\usepackage{multirow}

\usepackage{natbib}
\bibpunct{(}{)}{;}{a}{}{,}

\usepackage[switch]{lineno}

\newcommand{\til}{\raise.17ex\hbox{$\scriptstyle\sim$}}
\begin{document}

%\linenumbers

%\renewcommand{\baselinestretch}{2.0}\normalsize

%\title{Optical follow-up program of IceCube multiplets - testing for soft relativistic jets in Core-collapse Supernovae}
\title{Searching for soft relativistic jets in Core-collapse Supernovae with the IceCube Optical Follow-up Program}

\author{
  R.~Abbasi\inst{\ref{inst28}} 
  \and Y.~Abdou\inst{\ref{inst22}} 
  \and T.~Abu-Zayyad\inst{\ref{inst33}} 
  \and M.~Ackermann\inst{\ref{inst39}} 
  \and J.~Adams\inst{\ref{inst16}} 
  \and J.~A.~Aguilar\inst{\ref{inst28}} 
  \and M.~Ahlers\inst{\ref{inst32}} 
  \and M.~M.~Allen\inst{\ref{inst36}} 
  \and D.~Altmann\inst{\ref{inst1}} 
  \and K.~Andeen\inst{\ref{inst28},\ref{a}} 
  \and J.~Auffenberg\inst{\ref{inst38}} 
  \and X.~Bai\inst{\ref{inst31},\ref{b}} 
  \and M.~Baker\inst{\ref{inst28}} 
  \and S.~W.~Barwick\inst{\ref{inst24}} 
  \and R.~Bay\inst{\ref{inst7}} 
  \and J.~L.~Bazo~Alba\inst{\ref{inst39}} 
  \and K.~Beattie\inst{\ref{inst8}} 
  \and J.~J.~Beatty\inst{\ref{inst18},\ref{inst19}} 
  \and S.~Bechet\inst{\ref{inst13}} 
  \and J.~K.~Becker\inst{\ref{inst10}} 
  \and K.-H.~Becker\inst{\ref{inst38}} 
  \and M.~L.~Benabderrahmane\inst{\ref{inst39}} 
  \and S.~BenZvi\inst{\ref{inst28}}
  \and J.~Berdermann\inst{\ref{inst39}} 
  \and P.~Berghaus\inst{\ref{inst31}} 
  \and D.~Berley\inst{\ref{inst17}} 
  \and E.~Bernardini\inst{\ref{inst39}}
  \and D.~Bertrand\inst{\ref{inst13}} 
  \and D.~Z.~Besson\inst{\ref{inst26}} 
  \and D.~Bindig\inst{\ref{inst38}} 
  \and M.~Bissok\inst{\ref{inst1}}
  \and E.~Blaufuss\inst{\ref{inst17}}
  \and J.~Blumenthal\inst{\ref{inst1}} 
  \and D.~J.~Boersma\inst{\ref{inst1}} 
  \and C.~Bohm\inst{\ref{inst34}} 
  \and D.~Bose\inst{\ref{inst14}} 
  \and S.~B\"{o}ser\inst{\ref{inst11}} 
  \and O.~Botner\inst{\ref{inst37}}
  \and A.~M.~Brown\inst{\ref{inst16}}
  \and S.~Buitink\inst{\ref{inst14}}
  \and K.~S.~Caballero-Mora\inst{\ref{inst36}} 
  \and M.~Carson\inst{\ref{inst22}} 
  \and D.~Chirkin\inst{\ref{inst28}} 
  \and B.~Christy\inst{\ref{inst17}} 
  \and F.~Clevermann\inst{\ref{inst20}} 
  \and S.~Cohen\inst{\ref{inst25}} 
  \and C.~Colnard\inst{\ref{inst23}} 
  \and D.~F.~Cowen\inst{\ref{inst36},\ref{inst35}} 
  \and A.~H.~Cruz Silva\inst{\ref{inst39}}
  \and M.~V.~D'Agostino\inst{\ref{inst7}} 
  \and M.~Danninger\inst{\ref{inst34}} 
  \and J.~Daughhetee\inst{\ref{inst5}} 
  \and J.~C.~Davis\inst{\ref{inst18}} 
  \and C.~De Clercq\inst{\ref{inst14}} 
  \and T.~Degner\inst{\ref{inst11}}
  \and L.~Demir\"{o}rs\inst{\ref{inst25}} 
  \and F.~Descamps\inst{\ref{inst22}} 
  \and P.~Desiati\inst{\ref{inst28}} 
  \and G.~de Vries-Uiterweerd\inst{\ref{inst22}} 
  \and T.~DeYoung\inst{\ref{inst36}} 
  \and J.~C.~D\'{\i}az-V\'{e}lez\inst{\ref{inst28}} 
  \and M.~Dierckxsens\inst{\ref{inst13}} 
  \and J.~Dreyer\inst{\ref{inst10}} 
  \and J.~P.~Dumm\inst{\ref{inst28}} 
  \and M.~Dunkman\inst{\ref{inst36}} 
  \and J.~Eisch\inst{\ref{inst28}} 
  \and R.~W.~Ellsworth\inst{\ref{inst17}} 
  \and O.~Engdeg\aa{}rd\inst{\ref{inst37}} 
  \and S.~Euler\inst{\ref{inst1}} 
  \and P.~A.~Evenson\inst{\ref{inst31}} 
  \and O.~Fadiran\inst{\ref{inst28}} 
  \and A.~R.~Fazely\inst{\ref{inst6}} 
  \and A.~Fedynitch\inst{\ref{inst10}} 
  \and J.~Feintzeig\inst{\ref{inst28}} 
  \and T.~Feusels\inst{\ref{inst22}} 
  \and K.~Filimonov\inst{\ref{inst7}}
  \and C.~Finley\inst{\ref{inst34}} 
  \and T.~Fischer-Wasels\inst{\ref{inst38}} 
  \and B.~D.~Fox\inst{\ref{inst36}} 
  \and A.~Franckowiak\inst{\ref{inst11}}\thanks{franckowiak@physik.uni-bonn.de} 
  \and R.~Franke\inst{\ref{inst39}} 
  \and T.~K.~Gaisser\inst{\ref{inst31}} 
  \and J.~Gallagher\inst{\ref{inst27}} 
  \and L.~Gerhardt\inst{\ref{inst8}\ref{inst7}} 
  \and L.~Gladstone\inst{\ref{inst28}} 
  \and T.~Gl\"{u}senkamp\inst{\ref{inst39}} 
  \and A.~Goldschmidt\inst{\ref{inst8}} 
  \and J.~A.~Goodman\inst{\ref{inst17}} 
  \and D.~G\'{o}ra\inst{\ref{inst39}} 
  \and D.~Grant\inst{\ref{inst21}} 
  \and T.~Griesel\inst{\ref{inst29}} 
  \and A.~Gro\ss{}\inst{\ref{inst16},\ref{inst23}}
  \and S.~Grullon\inst{\ref{inst28}} 
  \and M.~Gurtner\inst{\ref{inst38}} 
  \and C.~Ha\inst{\ref{inst36}}
  \and A.~Haj~Ismail\inst{\ref{inst22}} 
  \and A.~Hallgren\inst{\ref{inst37}}
  \and F.~Halzen\inst{\ref{inst28}} 
  \and K.~Han\inst{\ref{inst39}} 
  \and K.~Hanson\inst{\ref{inst13},\ref{inst28}}
  \and D.~Heinen\inst{\ref{inst1}} 
  \and K.~Helbing\inst{\ref{inst38}} 
  \and R.~Hellauer\inst{\ref{inst17}} 
  \and P.~Herquet\inst{\ref{inst30}} 
  \and S.~Hickford\inst{\ref{inst16}} 
  \and G.~C.~Hill\inst{\ref{inst28}}
  \and K.~D.~Hoffman\inst{\ref{inst17}} 
  \and B.~Hoffmann\inst{\ref{inst1}}
  \and A.~Homeier\inst{\ref{inst11}} 
  \and K.~Hoshina\inst{\ref{inst28}}
  \and W.~Huelsnitz\inst{\ref{inst17},\ref{c}} 
  \and J.-P.~H\"{u}l\ss{}\inst{\ref{inst1}} 
  \and P.~O.~Hulth\inst{\ref{inst34}} 
  \and K.~Hultqvist\inst{\ref{inst34}} 
  \and S.~Hussain\inst{\ref{inst31}} 
  \and A.~Ishihara\inst{\ref{inst15}} 
  \and E.~Jacobi\inst{\ref{inst39}}
  \and J.~Jacobsen\inst{\ref{inst28}} 
  \and G.~S.~Japaridze\inst{\ref{inst4}}
  \and H.~Johansson\inst{\ref{inst34}} 
  \and K.-H.~Kampert\inst{\ref{inst38}} 
  \and A.~Kappes\inst{\ref{inst9}} 
  \and T.~Karg\inst{\ref{inst38}} 
  \and A.~Karle\inst{\ref{inst28}} 
  \and P.~Kenny\inst{\ref{inst26}} 
  \and J.~Kiryluk\inst{\ref{inst8},\ref{inst7}} 
  \and F.~Kislat\inst{\ref{inst39}} 
  \and S.~R.~Klein\inst{\ref{inst8},\ref{inst7}} 
  \and J.-H.~K\"{o}hne\inst{\ref{inst20}} 
  \and G.~Kohnen\inst{\ref{inst30}} 
  \and H.~Kolanoski\inst{\ref{inst9}} 
  \and L.~K\"{o}pke\inst{\ref{inst29}} 
  \and S.~Kopper\inst{\ref{inst38}} 
  \and D.~J.~Koskinen\inst{\ref{inst36}} 
  \and M.~Kowalski\inst{\ref{inst11}} 
  \and T.~Kowarik\inst{\ref{inst29}} 
  \and M.~Krasberg\inst{\ref{inst28}} 
  \and G.~Kroll\inst{\ref{inst29}} 
  \and N.~Kurahashi\inst{\ref{inst28}} 
  \and T.~Kuwabara\inst{\ref{inst31}} 
  \and M.~Labare\inst{\ref{inst14}} 
  \and K.~Laihem\inst{\ref{inst1}} 
  \and H.~Landsman\inst{\ref{inst28}} 
  \and M.~J.~Larson\inst{\ref{inst36}} 
  \and R.~Lauer\inst{\ref{inst39}} 
  \and J.~L\"{u}nemann\inst{\ref{inst29}} 
  \and J.~Madsen\inst{\ref{inst33}} 
  \and A.~Marotta\inst{\ref{inst13}} 
  \and R.~Maruyama\inst{\ref{inst28}} 
  \and K.~Mase\inst{\ref{inst15}} 
  \and H.~S.~Matis\inst{\ref{inst8}}
  \and K.~Meagher\inst{\ref{inst17}} 
  \and M.~Merck\inst{\ref{inst28}}
  \and P.~M\'{e}sz\'{a}ros\inst{\ref{inst35},\ref{inst36}} 
  \and T.~Meures\inst{\ref{inst13}} 
  \and S.~Miarecki\inst{\ref{inst8},\ref{inst7}} 
  \and E.~Middell\inst{\ref{inst39}} 
  \and N.~Milke\inst{\ref{inst20}} 
  \and J.~Miller\inst{\ref{inst37}} 
  \and T.~Montaruli\inst{\ref{inst28},\ref{d}} 
  \and R.~Morse\inst{\ref{inst28}} 
  \and S.~M.~Movit\inst{\ref{inst35}} 
  \and R.~Nahnhauer\inst{\ref{inst39}} 
  \and J.~W.~Nam\inst{\ref{inst24}} 
  \and U.~Naumann\inst{\ref{inst38}}
  \and D.~R.~Nygren\inst{\ref{inst8}} 
  \and S.~Odrowski\inst{\ref{inst23}} 
  \and A.~Olivas\inst{\ref{inst17}} 
  \and M.~Olivo\inst{\ref{inst10}} 
  \and A.~O'Murchadha\inst{\ref{inst28}}
  \and S.~Panknin\inst{\ref{inst11}} 
  \and L.~Paul\inst{\ref{inst1}} 
  \and C.~P\'{e}rez de los Heros\inst{\ref{inst37}} 
  \and J.~Petrovic\inst{\ref{inst13}} 
  \and A.~Piegsa\inst{\ref{inst29}}
  \and D.~Pieloth\inst{\ref{inst20}} 
  \and R.~Porrata\inst{\ref{inst7}}
  \and J.~Posselt\inst{\ref{inst38}} 
  \and P.~B.~Price\inst{\ref{inst7}} 
  \and G.~T.~Przybylski\inst{\ref{inst8}}
  \and K.~Rawlins\inst{\ref{inst3}}
  \and P.~Redl\inst{\ref{inst17}} 
  \and E.~Resconi\inst{\ref{inst23},\ref{e}} 
  \and W.~Rhode\inst{\ref{inst20}} 
  \and M.~Ribordy\inst{\ref{inst25}} 
  \and M.~Richman\inst{\ref{inst17}} 
  \and J.~P.~Rodrigues\inst{\ref{inst28}} 
  \and F.~Rothmaier\inst{\ref{inst29}}
  \and C.~Rott\inst{\ref{inst18}} 
  \and T.~Ruhe\inst{\ref{inst20}} 
  \and D.~Rutledge\inst{\ref{inst36}} 
  \and B.~Ruzybayev\inst{\ref{inst31}} 
  \and D.~Ryckbosch\inst{\ref{inst22}} 
  \and H.-G.~Sander\inst{\ref{inst29}}
  \and M.~Santander\inst{\ref{inst28}} 
  \and S.~Sarkar\inst{\ref{inst32}} 
  \and K.~Schatto\inst{\ref{inst29}} 
  \and T.~Schmidt\inst{\ref{inst17}} 
  \and A.~Sch\"{o}nwald\inst{\ref{inst39}} 
  \and A.~Schukraft\inst{\ref{inst1}} 
  \and A.~Schultes\inst{\ref{inst38}} 
  \and O.~Schulz\inst{\ref{inst23},\ref{f}} 
  \and M.~Schunck\inst{\ref{inst1}} 
  \and D.~Seckel\inst{\ref{inst31}} 
  \and B.~Semburg\inst{\ref{inst38}} 
  \and S.~H.~Seo\inst{\ref{inst34}} 
  \and Y.~Sestayo\inst{\ref{inst23}} 
  \and S.~Seunarine\inst{\ref{inst12}} 
  \and A.~Silvestri\inst{\ref{inst24}} 
  \and G.~M.~Spiczak\inst{\ref{inst33}}
  \and C.~Spiering\inst{\ref{inst39}} 
  \and M.~Stamatikos\inst{\ref{inst18},\ref{g}} 
  \and T.~Stanev\inst{\ref{inst31}} 
  \and T.~Stezelberger\inst{\ref{inst8}} 
  \and R.~G.~Stokstad\inst{\ref{inst8}} 
  \and A.~St\"{o}ssl\inst{\ref{inst39}} 
  \and E.~A.~Strahler\inst{\ref{inst14}}
  \and R.~Str\"{o}m\inst{\ref{inst37}}
  \and M.~St\"{u}er\inst{\ref{inst11}}
  \and G.~W.~Sullivan\inst{\ref{inst17}} 
  \and Q.~Swillens\inst{\ref{inst13}} 
  \and H.~Taavola\inst{\ref{inst37}}
  \and I.~Taboada\inst{\ref{inst5}} 
  \and A.~Tamburro\inst{\ref{inst33}}
  \and S.~Ter-Antonyan\inst{\ref{inst6}}
  \and S.~Tilav\inst{\ref{inst31}} 
  \and P.~A.~Toale\inst{\ref{inst2}} 
  \and S.~Toscano\inst{\ref{inst28}} 
  \and D.~Tosi\inst{\ref{inst39}} 
  \and N.~van Eijndhoven\inst{\ref{inst14}} 
  \and J.~Vandenbroucke\inst{\ref{inst7}} 
  \and A.~Van Overloop\inst{\ref{inst22}}
  \and J.~van~Santen\inst{\ref{inst28}} 
  \and M.~Vehring\inst{\ref{inst1}}
  \and M.~Voge\inst{\ref{inst11}} 
  \and C.~Walck\inst{\ref{inst34}} 
  \and T.~Waldenmaier\inst{\ref{inst9}} 
  \and M.~Wallraff\inst{\ref{inst1}} 
  \and M.~Walter\inst{\ref{inst39}} 
  \and Ch.~Weaver\inst{\ref{inst28}}
  \and C.~Wendt\inst{\ref{inst28}}
  \and S.~Westerhoff\inst{\ref{inst28}}
  \and N.~Whitehorn\inst{\ref{inst28}} 
  \and K.~Wiebe\inst{\ref{inst29}} 
  \and C.~H.~Wiebusch\inst{\ref{inst1}} 
  \and D.~R.~Williams\inst{\ref{inst2}}
  \and R.~Wischnewski\inst{\ref{inst39}} 
  \and H.~Wissing\inst{\ref{inst17}} 
  \and M.~Wolf\inst{\ref{inst23}} 
  \and T.~R.~Wood\inst{\ref{inst21}}
  \and K.~Woschnagg\inst{\ref{inst7}} 
  \and C.~Xu\inst{\ref{inst31}} 
  \and D.~L.~Xu\inst{\ref{inst2}} 
  \and X.~W.~Xu\inst{\ref{inst6}} 
  \and J.~P.~Yanez\inst{\ref{inst39}}
  \and G.~Yodh\inst{\ref{inst24}} 
  \and S.~Yoshida\inst{\ref{inst15}} 
  \and P.~Zarzhitsky\inst{\ref{inst2}}
  \and M.~Zoll\inst{\ref{inst34}} (the IceCube Collaboration)
  \and and C.~W.~Akerlof\inst{\ref{inst50}}  
  \and S.~B.~Pandey\inst{\ref{instPandey}} 
  \and F.~Yuan\inst{\ref{inst51}}
  \and W.~Zheng\inst{\ref{inst50}} (the ROTSE Collaboration)
}
%\author{A. Franckowiak\altaffilmark{1}, M.Kowalski\altaffilmark{1} for the IceCube Collaboration}
%\affil{Physikalisches Institut, Universit\"{a}t Bonn}

\institute{\tiny
  III. Physikalisches Institut, RWTH Aachen University, D-52056 Aachen, Germany\label{inst1} \and
  Dept. of Physics and Astronomy, University of Alabama, Tuscaloosa, AL 35487, USA\label{inst2} \and
  Dept. of Physics and Astronomy, University of Alaska Anchorage, 3211 Providence Dr., Anchorage, AK 99508, USA\label{inst3} \and
  University of Michigan, Randall Laboratory of Physics, 450 Church St., Ann Arbor, MI, 48109-1040\label{inst50} \and
  CTSPS, Clark-Atlanta University, Atlanta, GA 30314, USA\label{inst4} \and
  School of Physics and Center for Relativistic Astrophysics, Georgia Institute of Technology, Atlanta, GA 30332, USA\label{inst5} \and
  Dept. of Physics, Southern University, Baton Rouge, LA 70813, USA\label{inst6} \and
  Dept. of Physics, University of California, Berkeley, CA 94720, USA\label{inst7} \and
  Lawrence Berkeley National Laboratory, Berkeley, CA 94720, USA\label{inst8} \and
  Institut für Physik, Humboldt-Universit\"{a}t zu Berlin, D-12489 Berlin, Germany\label{inst9} \and
  Fakult\"{a}t für Physik \& Astronomie, Ruhr-Universit\"{a}t Bochum, D-44780 Bochum, Germany\label{inst10} \and
  Physikalisches Institut, Universit\"{a}t Bonn, Nussallee 12, D-53115 Bonn, Germany\label{inst11} \and
  Dept. of Physics, University of the West Indies, Cave Hill Campus, Bridgetown BB11000, Barbados\label{inst12} \and
  Universit\'{e} Libre de Bruxelles, Science Faculty CP230, B-1050 Brussels, Belgium\label{inst13} \and
  Vrije Universiteit Brussel, Dienst ELEM, B-1050 Brussels, Belgium\label{inst14} \and
  Dept. of Physics, Chiba University, Chiba 263-8522, Japan\label{inst15} \and
  Dept. of Physics and Astronomy, University of Canterbury, Private Bag 4800, Christchurch, New Zealand\label{inst16} \and
  Dept. of Physics, University of Maryland, College Park, MD 20742, USA\label{inst17} \and
  Dept. of Physics and Center for Cosmology and Astro-Particle Physics, Ohio State University, Columbus, OH 43210, USA\label{inst18} \and
  Dept. of Astronomy, Ohio State University, Columbus, OH 43210, USA\label{inst19} \and
  Dept. of Physics, TU Dortmund University, D-44221 Dortmund, Germany\label{inst20} \and
  Dept. of Physics, University of Alberta, Edmonton, Alberta, Canada T6G 2G7\label{inst21} \and
  Dept. of Physics and Astronomy, University of Gent, B-9000 Gent, Belgium\label{inst22} \and
  Max-Planck-Institut für Kernphysik, D-69177 Heidelberg, Germany\label{inst23} \and
  Dept. of Physics and Astronomy, University of California, Irvine, CA 92697, USA\label{inst24} \and
  Laboratory for High Energy Physics, \'{E}cole Polytechnique F\'{e}d\'{e}rale, CH-1015 Lausanne, Switzerland\label{inst25} \and
  Dept. of Physics and Astronomy, University of Kansas, Lawrence, KS 66045, USA\label{inst26} \and
  Dept. of Astronomy, University of Wisconsin, Madison, WI 53706, USA\label{inst27} \and
  Dept. of Physics, University of Wisconsin, Madison, WI 53706, USA\label{inst28} \and
  Institute of Physics, University of Mainz, Staudinger Weg 7, D-55099 Mainz, Germany\label{inst29} \and
  Universit\'{e} de Mons, 7000 Mons, Belgium\label{inst30} \and
  Aryabhatta Research Institute of Observational  Sciences (ARIES), Nainital, India\label{instPandey} \and
  Bartol Research Institute and Department of Physics and Astronomy, University of Delaware, Newark, DE 19716, USA\label{inst31} \and
  Dept. of Physics, University of Oxford, 1 Keble Road, Oxford OX1 3NP, UK\label{inst32} \and
  Dept. of Physics, University of Wisconsin, River Falls, WI 54022, USA\label{inst33} \and
  Oskar Klein Centre and Dept. of Physics, Stockholm University, SE-10691 Stockholm, Sweden\label{inst34} \and
  Dept. of Astronomy and Astrophysics, Pennsylvania State University, University Park, PA 16802, USA\label{inst35} \and
  Dept. of Physics, Pennsylvania State University, University Park, PA 16802, USA\label{inst36} \and
  Dept. of Physics and Astronomy, Uppsala University, Box 516, S-75120 Uppsala, Sweden\label{inst37} \and
  Research School of Astronomy and Astrophysics, The Australian National University, Cotter Road, Weston Creek, ACT 2611, Australia\label{inst51} \and
  Dept. of Physics, University of Wuppertal, D-42119 Wuppertal, Germany\label{inst38} \and
  DESY, D-15735 Zeuthen, Germany \label{inst39} \and
  now at Dept. of Physics and Astronomy, Rutgers University, Piscataway, NJ 08854, USA\label{a} \and
  now at Physics Department, South Dakota School of Mines and Technology, Rapid City, SD 57701, USA\label{b} \and
  Los Alamos National Laboratory, Los Alamos, NM 87545, USA\label{c} \and
  also Sezione INFN, Dipartimento di Fisica, I-70126, Bari, Italy\label{d} \and
  now at T.U. Munich, 85748 Garching \& Friedrich-Alexander Universit\"{a}t Erlangen-N\"{u}rnberg, 91058 Erlangen, Germany\label{e} \and
  now at T.U. Munich, 85748 Garching, Germany\label{f} \and
  NASA Goddard Space Flight Center, Greenbelt, MD 20771, USA\label{g} 
}

\abstract
   {Transient neutrino sources such as Gamma-Ray Bursts (GRBs) and Supernovae (SNe) are hypothesized to 
emit bursts of high-energy neutrinos on a time-scale of $\lesssim 100$\,s. While GRB neutrinos would be produced in high relativistic jets, core-collapse SNe might host soft-relativistic jets, which become stalled in the outer layers of the progenitor star leading to an efficient production of high-energy neutrinos.}
   {To increase the sensitivity to these neutrinos and identify their sources, a low-threshold optical follow-up program for neutrino multiplets detected with the IceCube observatory has been implemented.}
   {If a neutrino multiplet, i.e. two or more neutrinos from the same direction within $100$\,s, is found by IceCube
a trigger is sent to the Robotic Optical Transient Search Experiment, ROTSE. The 4 ROTSE telescopes immediately start an observation program of the corresponding region 
of the sky in order to detect an optical counterpart to the neutrino events.}
   {No statistically significant excess in the rate of neutrino multiplets has been observed and furthermore no coincidence with an optical counterpart was found.}
   {The search allows, for the first time, to set stringent limits on current models predicting a high-energy neutrino flux from soft relativistic hadronic jets in core-collapse SNe. We conclude that a sub-population of SNe with typical Lorentz boost factor and jet energy of $10$ and $3\times10^{51}$\,erg, respectively, does not exceed $4.2\%$ at $90\%$ confidence.
   %less than $7.8\%$ of all SNe have a jet with a Lorentz boost factor of $\Gamma=10$ and a typical GRb jet energy of $E_{\rm{jet}}=3\times10^{51}$\,erg.
}

\keywords{neutrinos -- supernovae -- gamma-ray bursts}

\titlerunning{IceCube Optical Follow-up Program}
\authorrunning{Abbasi et al.}
\maketitle

\section{Introduction}
%\begin{linenumbers}
High-energy astrophysical neutrinos are produced in interactions of charged cosmic
rays with ambient photon or baryonic fields (for reviews see \citet{Anchordoqui:2009nf,Chiarusi:2009ng,BeckerReview,Lipari:2006uw}). Acceleration of these cosmic rays to very high energies takes place in astrophysical shocks. 
%High-energy astrophysical neutrinos are produced in proton interactions of charged cosmic
%rays with ambient photon or baryonic fields, when the protons are 
%accelerated to very high energies in astrophysical shocks (for reviews see \citet{Anchordoqui:2009nf,Chiarusi:2009ng,BeckerReview,Lipari:2006uw}). 
Neutrinos escape the acceleration region and propagate through space %basically unaffected
without interaction, while nuclei are deflected in magnetic fields and
no longer point back to their source (for energies below \til$10^{20}$\,eV). Unlike gamma-rays, neutrinos are solely produced in hadronic processes and would therefore reveal the sources of charged cosmic rays.
Gamma-ray bursts (see \citet{GRBReviewMeszaros2001} and \citet{GRBReviewZhang2003} for reviews) could provide the environment and the required energy to explain the production of the highest energy cosmic-rays~\citep{Waxman95}. %and hence are a plausible candidate.
Recent observations indicate a connection between long GRBs (duration $\gtrsim 2$\,s) and core-collapse supernovae (CCSNe). In several cases a gamma-ray burst or X-ray flash has been observed in coincidence with an optical SN light curve implying a common physical origin: a massive stellar explosion (see~\citet{Woosley:2006fn} for a review).
%Coincidences were observed for SN1998bw/GRB980425~\citep{Galama:1998ea}, SN2003lw/GRB031203~\citep{:2004sg}, SN2003dh/GRB030329~\citep{Hjorth:2003jt}, SN2006aj/XRF060218~\citep{Pian:2006pr}, SN2008d/XRF080109~\citep{Soderberg:2008uh} and SN2010bh/GRB100316D~\citep{Starling:2010ed}.
%In some cases, spectroscopic observation of bumps observed during the late decline of GRB afterglows has revealed the presence of SN features.
Furthermore, GRBs and CCSNe were found to release a comparable amount of kinetic energy. 
According to the collapsar model \citep{MacFadyen:1998vz,Paczynski:1997yg,Woosley:1993wj}, long GRBs have their origin in 
the collapse of a massive, rapidly rotating star into a black hole surrounded by an accretion disk. Relativistic jets 
with Lorentz boost factors of $100-1000$ form along the stellar axis.
The GRB-SN connection gives rise to the idea that GRBs and SNe 
might have the jet signature in common and a certain fraction of core-collapse SNe might host soft relativistic jets with Lorentz boost factors around $5$. The Lorentz boost factor of the jet might be determined by features of the progenitor star, such as its rotation. Compared to jets in GRBs, SN jets are suggested to be equally energetic but more baryon-rich, hence they are only mildly relativistic.
Such soft relativistic jets would become stalled in the outer layers of the progenitor star, leading to essentially full absorption of
the electromagnetic radiation emitted by the jet and, at the same time, an efficient production
of high-energy neutrinos \citep{Razzaque:2005bh,AndoBeacom}. This motivates a search for neutrino emission, as neutrinos would be able to escape from within the star.\\
The IceCube neutrino detector, located at the geographic South Pole, is built to detect high-energy astrophysical neutrinos~\citep{IceCube}.
So far GRB neutrino searches have been performed offline on AMANDA~\citep{GRBAmanda,Achterberg:2007qy} and IceCube~\citep{GRBIC,Abbasi:2009kq,Abbasi:2009ig} data,
triggered by gamma-ray satellite 
detections. %Location and timing information 
Time and direction information provided by gamma-ray satellites
allow an almost background free search. An untriggered search was applied to AMANDA data~\citep{Achterberg:2007qy}, which scanned the data for a clustering of neutrinos in time. Furthermore, a dedicated search for a neutrino signal in coincidence with the observed X-ray flash of SN~2008D has been conducted by IceCube~\citep{Abbasi:2011wh} in order to test the soft jet scenario for CCSNe. 
Neither the GRB nor the SN neutrino search led to a detection yet, but 
%Both, GRB and SN neutrino searches, did not lead to a detection 
set upper
limits on the neutrino flux.\\
Early SN detections, as in the case of SN~2008D, are very rare since X-ray telescopes have a limited field of view (FoV).
However, neutrino telescopes cover half of the sky at any time. %the whole sky at any time. 
If neutrinos produced in soft relativistic SN jets are detected in real time, they can be used to trigger follow-up observations \citep{Kowalski:2007xb,Ageron:2011pe}. This is realized with the optical follow-up program presented here.
Complementary to the offline searches, the optical follow-up program is an 
online search independent of satellite detections. It is sensitive to transient objects, including those which are either gamma-ray dark or
not detected by gamma-ray satellites. In addition to a gain in sensitivity, the optical observations may allow
the identification of the transient neutrino source, be it a SN, GRB or any other 
transient phenomenon producing an optical signal. Hence, it enables us to test 
the plausible hypothesis of a soft relativistic SN jet
and sheds light on the connection between GRBs, SNe and relativistic jets.\\
%If high-energy neutrino data from many supernovae could be collected,
%the distribution of the jet Lorentz boost factor could be obtained, providing
%important insight into the SN–GRB connection.\\
In order to implement the optical follow-up program an online neutrino event selection was developed for IceCube. The data are processed online by a computer farm at the South Pole. A dedicated trigger selects neutrino burst candidates and the directional information is transferred to the four ROTSE telescopes, which start the follow-up immediately and continue observations for several nights.
The obtained optical data are searched for a transient counterpart.\\
In the following, we present the optical follow-up program starting with section~\ref{sec:neutrinoFlux}, which briefly describes the expected neutrino emission according to the soft SN jet model.
Section~\ref{sec:NeutrinoDetection} outlines the IceCube component of the program
while section~\ref{sec:OpticalCounterpart} focuses on the search for the optical counterpart. 
Finally, we discuss systematic uncertainties in section~\ref{sec:SysErrors} and show our results from the first year of data taking
in section~\ref{sec:Results} with a focus on the SN soft jet model. We present a first limit on the
hadronic jet production in CCSNe and conclude with a summary and outlook to future extensions of the program.
%\end{linenumbers}

\section{SN neutrino flux}
\label{sec:neutrinoFlux}
\begin{figure}[t]
\centering
\includegraphics[angle=0,width=9cm, trim=0cm 0cm 0cm 0cm, clip=true]{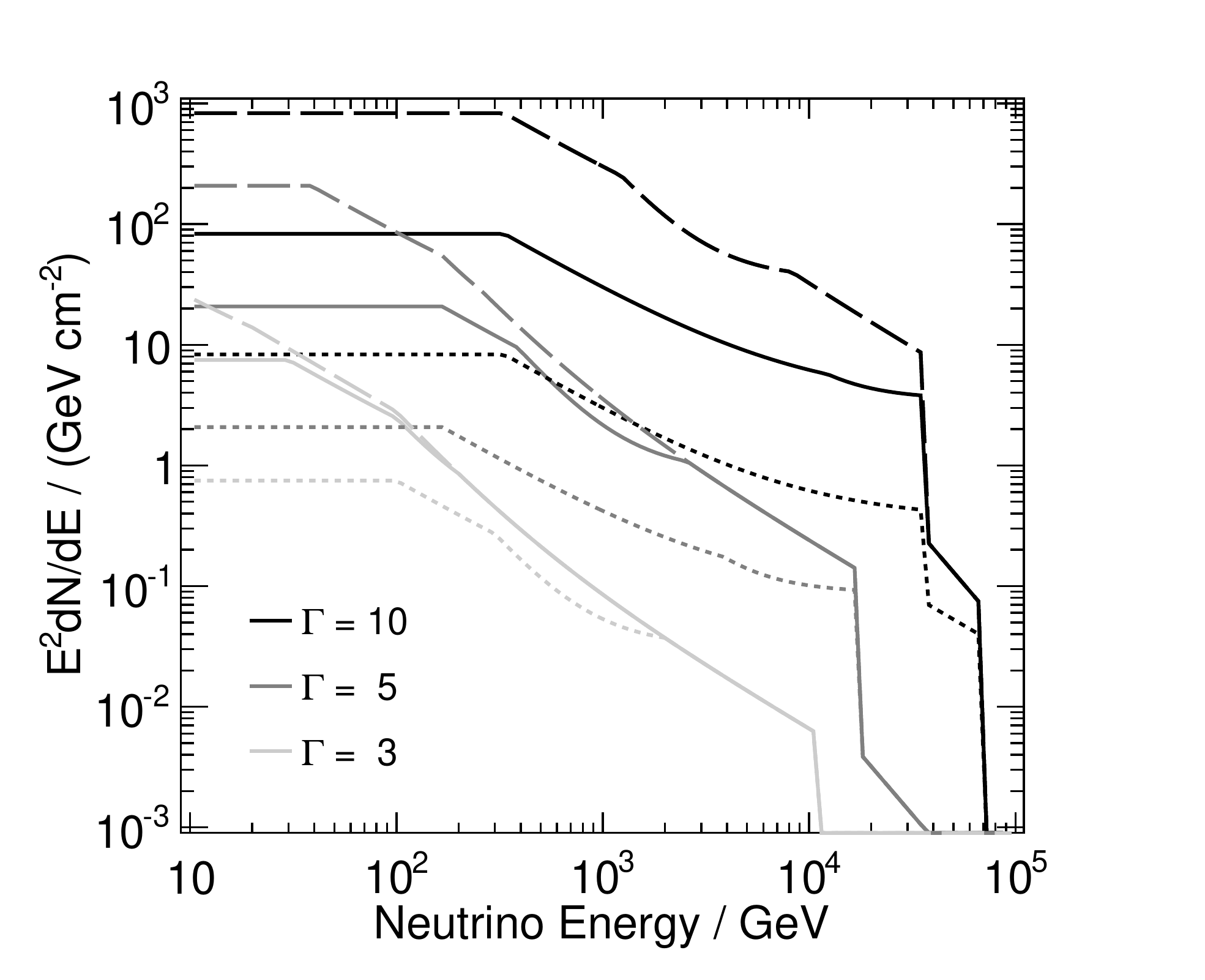}
\caption{SN neutrino spectrum according to AB05 for one SN at distance $10$\,Mpc with the jet pointing to us. Different shades of gray indicate different Lorentz boost factors. Solid lines: $E_{\rm{jet}}=3\times10^{51}$\,erg. Dotted line: $E_{\rm{jet}}=0.3\times10^{51}$\,erg. Dashed line: $E_{\rm{jet}}=30\times10^{51}$\,erg.}
\label{pic:ABFlux}
\end{figure}

%\begin{linenumbers}
Motivated by the GRB-SN connection \citet{Razzaque:2005bh} proposed a model for high-energy neutrino production in soft relativistic CCSN jets. 
If protons are accelerated in the jets through Fermi acceleration in internal shocks, proton-proton-collisions will produce kaons and pions. The initial formulation of the model only considered neutrino production through pion decay~\citep{Razzaque:2005bh}. It was extended by~\citet{AndoBeacom} (hereafter AB05), who included neutrino production from kaon decay yielding a harder and hence more easily detectable neutrino spectrum.
AB05 present the calculation of the neutrino spectrum
for a fixed Lorentz factor of $\Gamma_{0} = 3$ and a fixed jet energy of 
$E_{\rm{jet},0} = 3\times 10^{51}$\,erg. In order to test a broader parameter space we calculate the neutrino flux as a function of the Lorentz boost factor $\Gamma$, the jet energy $E_{\rm{jet}}$ and the density, $\rho$, of CCSN producing a jet. 
In the following, we derive the neutrino flux for one SN at distance $10$\,Mpc assuming it hosts a jet pointing at us following the calculation of AB05.\\
In the AB05 model, pions and kaons are produced with $20\%$ of the parent proton energy and follow the original
$E^{-2}$ energy spectrum originating from Fermi acceleration. %Radiative and hadronic cooling, i.e. energy loss, cause a steepening of the spectrum at higher energies. 
However, protons lose energy through synchrotron radiation and inverse Compton scattering (radiative cooling) and through $\pi p$ and $Kp$ processes (hadronic cooling)
causing a steepening of the spectrum at higher energies. Above a certain
break energy hadronic cooling becomes dominant and steepens the spectrum by a factor $E^{-1}$ 
while radiative cooling dominates above a second break energy resulting in a total suppression factor
of $E^{-2}$.\\
%Above a certain
%break energy $E_{\nu,\pi(K)}^{\rm{had}}$ hadronic cooling becomes dominant and steepens the spectrum by a factor $E^{-1}$ 
%while radiative cooling dominates above a second break energy $E_{\nu,\pi(K)}^{\rm{rad}}$ resulting in a total suppression factor
%of $E^{-2}$.  
The daughter neutrino carries $25\%$ of the pion energy or $50\%$ of the kaon energy and its
energy is related to the meson energy in the jet frame\footnote{Variables in the comoving frame, i.e. the frame of the jet, are denoted with a prime.} by
%\end{linenumbers}
\begin{equation}
 E_{\nu,\pi(K)} = \Gamma E_{\pi(K)}' /4(2).
\end{equation}
%\begin{equation}
% E_{\nu,\pi} = \Gamma E_{\pi}' /4 \textrm{ and }E_{\nu,K} = \Gamma E_{K}' /2.
%\end{equation}
%in the case of pions and 
%\begin{equation}
% E_{\nu,K} = \Gamma E_{K}' /2
%\end{equation}
%for kaons. \
%\begin{linenumbers}
AB05 assume a hadronic cooling break of $30$\,GeV for pions and $200$\,GeV for kaons. It depends 
%The hadronic cooling break depends 
strongly on the jet Lorentz factor
$\Gamma$ %(assuming $\Gamma = \theta^{-1}$, where $\theta$ is the jet opening angle) 
and the jet energy $E_{\rm{jet}}$:
%\end{linenumbers}
\begin{equation}
 E_{\nu,\pi(K)}^{\rm{had}} = \left(\frac{E_{\rm{jet}}}{E_{\rm{jet},0}} \right)^{-1} \left(  \frac{\Gamma}{\Gamma_{0}} \right)^{5} 30(200) \rm{\,GeV}.
\end{equation}
%\begin{equation}
% E_{\nu,\pi}^{(1)} = \left(\frac{E_{\textrm{jet}}}{E_{\textrm{jet},0}} \right)^{-1} \left(  \frac{\Gamma}{\Gamma_{0}} \right)^{5} 30 \textrm{~GeV}
%\end{equation}
%\begin{equation}
% E_{\nu,K}^{(1)} = \left(\frac{E_{\textrm{jet}}}{E_{\textrm{jet},0}} \right)^{-1} \left( \frac{\Gamma}{\Gamma_{0}} \right)^{5} 200 \textrm{~GeV}
%\end{equation}
%\begin{linenumbers}
The radiative cooling break ($0.1$\,GeV for pions and $20$\,GeV for kaons assumed by AB05) depends only on $\Gamma$:
%\end{linenumbers}
\begin{equation}
 E_{\nu,\pi(K)}^{\rm{rad}} =  \frac{\Gamma}{\Gamma_{0}} 0.1(20) \textrm{\,TeV}.
\end{equation}
%\begin{equation}
% E_{\nu,\pi}^{(2)} =  \frac{\Gamma}{\Gamma_{0}} 100 \textrm{~GeV}
%\end{equation}
%\begin{equation}
% E_{\nu,K}^{(2)} = \frac{\Gamma}{\Gamma_{0}} 20000 \textrm{~GeV}
%\end{equation}
%\begin{linenumbers}
Note that for certain parameter configurations the order of the two break energies can change.
Finally, the proton energy reaches its maximum at the photo-pion production threshold of 
$E_{p,max}' = 7 \times 10^{4}$\,GeV, where protons interact with the dense field of $4$\,keV thermalized synchrotron photons. Neutrinos produced in resulting $\Delta^+$-decays are not considered by AB05. The cut-off in the proton spectrum results in a cut-off in the neutrino spectrum at
%\end{linenumbers}
\begin{equation}
 E_{\nu,\pi(K)}^{\rm{cutoff}} =  \frac{\Gamma}{\Gamma_{0}} 10.5(21) \textrm{\,TeV}.
\end{equation}
%\begin{equation}
% E_{\nu,\pi}^{cutoff} =  \frac{\Gamma}{\Gamma_{0}} 10500 \textrm{~GeV}
%\end{equation}
%\begin{equation}
% E_{\nu,K}^{cutoff} = \frac{\Gamma}{\Gamma_{0}} 21000 \textrm{~GeV}
%\end{equation}
%\begin{linenumbers}
The normalization of the neutrino flux at 1\,GeV scales with the jet energy. 
%\end{linenumbers}
\begin{equation}
 F_{\nu,\pi(K),0} = \frac{E_{\rm{jet}}}{E_{\rm{jet},0}}  \left( \frac{\Gamma}{\Gamma_{0}} \right)^2  5 \times 10^{-5} (5 \times 10^{-2}) \textrm{\,GeV}^{-1}\textrm{cm}^{-2},
\end{equation}
%\begin{equation}
% F_{\nu,\pi,0} = \frac{E_{\textrm{jet}}}{E_{\textrm{jet},0}}  \left( \frac{\Gamma}{\Gamma_{0}} \right)^2  5 \cdot 10^{-5} \textrm{~GeV}^{-1}\textrm{cm}^{-2}
%\end{equation}
%\begin{equation}
% F_{\nu,K,0} =  \frac{E_{\textrm{jet}}}{E_{\textrm{jet},0}}  \left( \frac{\Gamma}{\Gamma_{0}} \right)^2  5 \cdot 10^{-2} \textrm{~GeV}^{-1}\textrm{cm}^{-2}
%\end{equation}
%\begin{linenumbers}
where the $\Gamma^2$ dependence is due to the assumed beaming with a jet opening angle of $\theta \propto 1/\Gamma$.
Note that the expected number of neutrinos does not simply scale with the jet energy because the
first break energy shifts with the jet energy. Depending on the choice of model parameters both the pion and kaon component of the spectrum can be hard ($\propto E^{-2}$) or soft ($\propto E^{-3}$) in the energy range, which IceCube is sensitive to (TeV to PeV). Figure~\ref{pic:ABFlux} illustrates the behavior of the total neutrino spectrum (sum of pion and kaon component) for different jet energies and Lorentz boost factors.
%Depending on the choice of model parameters the spectrum can be hard %($\propto E^{-2}$) or soft ($\propto E^{-3}$) % or $\propto E^{-4}$) 
%(\til$E^{-2}$) or soft (\til$E^{-3}$) 
%in the energy range, which IceCube is sensitive to (TeV to PeV).
%\end{linenumbers}

\section{IceCube}
\label{sec:NeutrinoDetection}
\begin{figure}[t]
\centering
\includegraphics[angle=0,width=9cm, trim=3cm 0cm 0cm 0cm, clip=true]{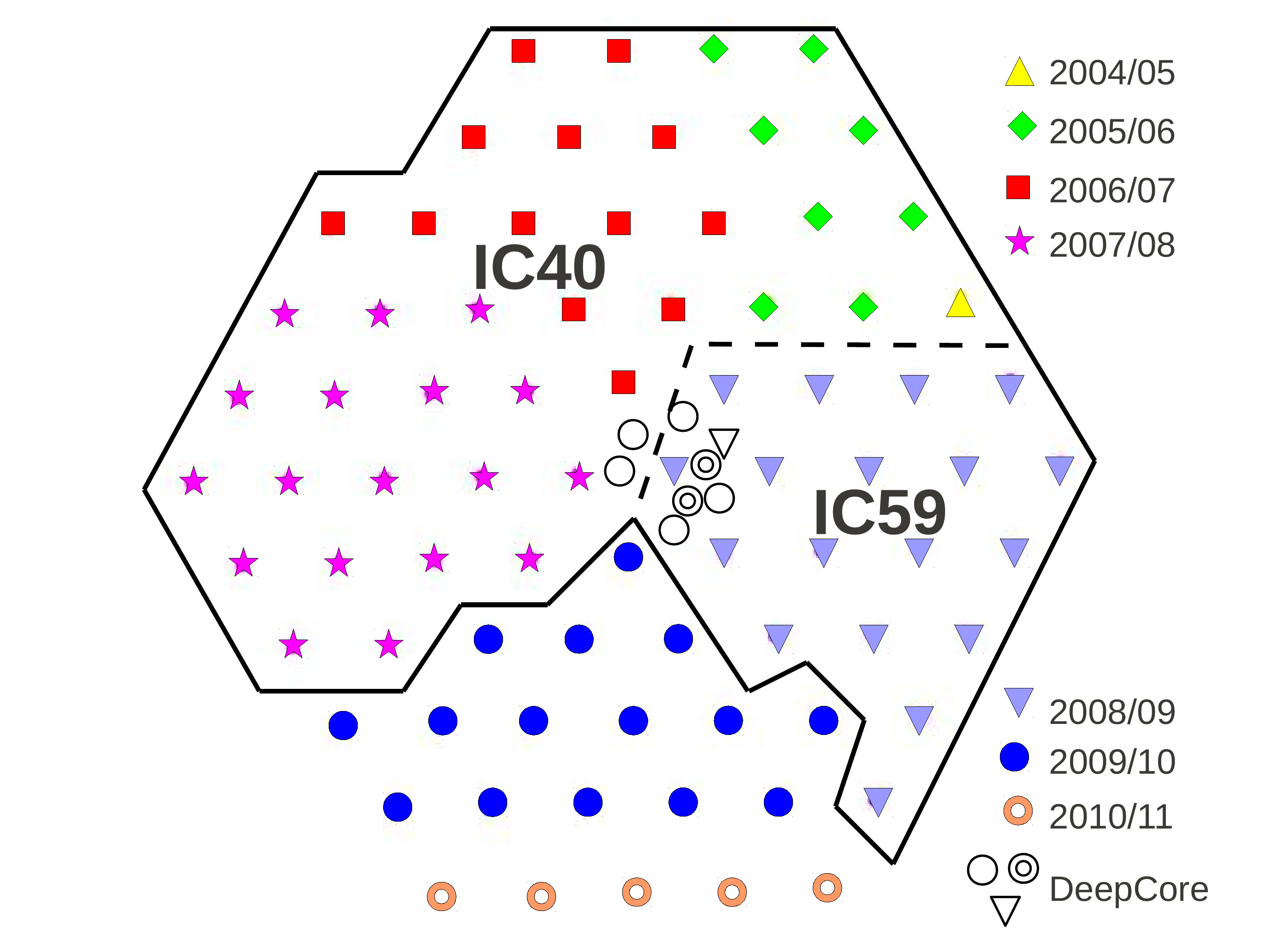}
\caption{The IceCube detector: The full detector consists of 86 strings with 60 DOMs attached to each string. Different colors/symbols indicate different deployment stages. %Yellow: Season 2004/05, first string. Green: Season 2005/06, 9 strings. Red: Season 2006/07, 22 strings. Magenta: Season 2007/08, 40 strings. Purple: Season 2008/09, 59 strings, including first DeepCore string. Blue: Season 2009/10, 79 strings, including 6 DeepCore strings. Orange: 86 strings, including 8 DeepCore strings. 
The solid black line encircles the IC59 configuration, while the dashed line indicates the smaller IC40 configuration.}
\label{pic:IceCube}
\end{figure}

%\begin{linenumbers}
The IceCube neutrino telescope has been under construction at the geographic South Pole 
since 2004 and was completed in the Antarctic summer of 2010/11. 
It is capable of detecting high energy neutrinos with energies of $\mathcal{O}(100)$\,GeV and is most sensitive to muon neutrinos with energies in the TeV range and above.
High-energy muon neutrinos undergoing charged current interactions in the ice or the underlying rock
produce muons in neutrino-nucleon interactions.
The muon travels in a direction close to that of the neutrino and emits Cherenkov light, which is detected by a three dimensional array of light sensors. 
%The deep clear Antarctic ice is instrumented with light sensors thus forming a Cherenkov particle detector.
For this purpose, a 
%It comprises a 
volume of $1$\,km$^3$ of clear ice in depths between $1450$ and $2450$\,m below the ice surface was instrumented 
with 5160 digital optical modules (DOMs) attached to 86 vertical strings %at a depth of $1450$\,m to 
%$2450$\,m 
\citep{IceCube}. %below the surface of the South Pole ice. 
Each DOM consists of a $25$\,cm diameter Hamamatsu photomultiplier tube (PMT) and supporting hardware inside a 
pressure glass sphere \citep{IceCubeDOM}. 
The detector consists of 78 strings arranged in a hexagonal shape with a string spacing of 125\,m 
and DOMs separated vertically by 17\,m, and 8 strings composing the low-energy extension DeepCore \citep{DeepCore}, a
densely spaced array in the bottom half of the detector. 
%78 of the strings are arranged in a hexagonal shape with a string spacing of 125\,m 
%and DOMs separated vertically by 17\,m, while 8 strings compose the low-energy extension DeepCore \citep{DeepCore}, a
%densely spaced array at the bottom half of the detector. 
%Six of the DeepCore strings use high
%quantum efficiency DOMs and have a dense spacing of about 70\,m horizontally and 7\,m vertically. The other two
%DeepCore strings are equipped with standard IceCube DOMs at 7\,m vertical spacing and have a
%horizontal spacing of 42\,m.
The observatory also includes a surface array, IceTop, to measure properties of extensive air showers
and study the composition and spectrum of cosmic rays \citep{Stanev:2009ce}.
Figure~\ref{pic:IceCube} shows a top view of the IceCube detector including DeepCore.
Different colors/symbols indicate different deployment stages. Before completion of the full detector, IceCube took data with the available number of strings.
The optical follow-up program has been fully operational since Dec. 16, 2008. Here, we present the analysis of the data taken from Dec. 16, 2008 to Dec. 31, 2009. 
Initially, 40 IceCube strings were taking data (yellow upward-pointing triangle, green diamonds, red squares and magenta stars in Fig.~\ref{pic:IceCube}). In May 2009, additional 19 strings were included (strings marked by purple downward-pointing triangles in Fig.~\ref{pic:IceCube} including the first DeepCore string marked by an open downward-pointing triangle). In the following, these deployment stages will be referred to as IC40 and IC59, respectively. \\
To suppress the background caused by PMT noise 
or radioactive decay in the glass, a total number of eight DOMs with coincident hits
%needs to register a coincident hit 
in a time window of $5$\,$\mu$s are required for a trigger to be formed. A coincident hit is registered when a single DOM and its
neighbor or next-to-neighbor DOM on the same string exceed their charge threshold of $0.25$\,photoelectrons
within a time window of $1$\,$\mu$s.
If the detector is triggered the information of all triggered DOMs within
a readout window starting $10$\,$\mu$s before the first hit and ending $10$\,$\mu$s after the last hit is read out and merged to an event. Overlapping readout windows are merged together. The waveform of the PMT 
is digitized and sent to the surface. The waveforms have a length of up to $6.4$\,$\mu$s and can contain multiple hits.
The total number of photoelectrons and their arrival times are extracted with an iterative 
%Bayesian-based 
unfolding algorithm.\\
The arrival time of the Cherenkov photons can 
be measured with an accuracy of \til$3.3$\,ns~\citep{IceCubeDOM}. Several hit DOMs allow a reconstruction of the direction of
muon-neutrinos with a precision of \til$1^{\circ}$. To reduce the contribution from noise, only hits within a $6$\,$\mu$s time window are used for the reconstruction (time window cleaning). This time window is defined as the window that contains the most hits during the event. Muons travel along a straight path through the detector while electrons or taus produce showers (cascades). %-like signatures. 
Only the muon signature allows an accurate reconstruction of the direction. Directional information is crucial
in order to provide coordinates for optical telescopes and we therefore consider only muon neutrino events.
%\end{linenumbers}

\subsection{Online System}
\label{subsec:OnlineSystem}
%\begin{linenumbers}
In order to rapidly trigger optical telescopes, the first online analysis of high-energy neutrino events in IceCube was implemented.
Unlike for the offline analyses, which are performed on an entire dataset (usually \til$1$\,year of data) with time consuming reconstructions on a large computer cluster,
the data are processed online by a computer cluster at the South Pole. During IC40 (IC59) data taking, the cluster consisted of approximately 60 (100) processing nodes. %\textbf{The nodes are contained in $2\times2$ core $2.4$\,GHz opteron as well as $4\times4$ core $3$\,GHz machines, each running a redhat enterprise 4 linux operating system.} 
%with $\sim60$~clients available during IC40 and $\sim100$~clients during IC59. 
The processing includes event reconstruction and basic event selection. 
Several filters and reconstruction algorithms run on the cluster (see section~\ref{subsection:eventSelection} for details on the filters and reconstructions).
During IC40 and IC59 events were read from the data acquisition and written to files with each file containing about $1$\,GB of data. The data volume of $1$\,GB corresponds to a data taking period of \til$5$\,min or \til$4\times10^5$ events for IC40. Due to larger event sizes the same volume relates to \til$2.5$\,min or \til$2.5\times10^5$ events in IC59. The files were then distributed to the processing nodes. Each file has to be processed by a single node. %Only one node can process each file. 
Although the processing time of \til$35$\,ms per event is fast, the processing time per file amounts to \til$4$\,h in IC40 and to \til$2.5$\,h in IC59. For technical reasons, an additional latency of $4$\,h was added to ensure correct ordering of the data after processing, which results in a total latency of $6.5-8$\,h\footnote{With the start of operations with 79 strings the parallel processing was upgraded, reducing the total latency to a few minutes.}. 
%However, the first year of data presented here was taken in the old mode with a latency of $6.5-8$\,h.\\
The processed and ordered data arrive on a dedicated machine (the analysis client), which is not part of the parallel processing. There, a sophisticated event selection is applied based on the reconstructed event parameters. A multiplet trigger selects candidates of neutrino clusters, which are coincident in time and space. No further reconstruction algorithms need to be applied allowing a very fast filtering of the events. If a multiplet is found, its directional information is transferred to Madison, Wisconsin, via the Iridium satellite network within $\lesssim 10$\,s. From there, the message is forwarded to the four ROTSE telescopes via the Internet through a TCP-socket connection for immediate follow-up observations. The stability and performance of the online system is constantly monitored in order to allow for a fast discovery of problems. % and to avoid sending fake alerts. 
To achieve this, test alerts are produced at a much higher rate (\til$100$ test alerts per day compared to 25 real alerts per year) by the same pipeline and are also sent to the North. Their rate and delay time distributions are monitored using an automatically generated web page. If the rate deviates significantly from the expected, a notification is issued. %the system responsibles are automatically notified.

%\end{linenumbers}

\subsection{Neutrino Data Set}
\label{subsec:NeutrinoDataSet}
%\begin{linenumbers}
We present the analysis of data taken from Dec. 16, 2008 to Dec. 31, 2009. 
This corresponds to a lifetime of 121\,days with IC40 and 186.4\,days with IC59. Dead time is predominantly caused by calibration and commissioning runs during and after the construction season. The downtime of the online system amounts to 6.8\% mainly caused by downtime of the satellite transmission system.\\
%The data set is strongly background dominated.
The background in a search for muon-neutrinos of astrophysical origin can be divided into two classes. One consists of atmospheric
muons, created in meson decays in cosmic ray air showers, entering the detector from above. The other background is 
atmospheric neutrinos which originate in the same meson decays in cosmic ray air showers. Both are measured with IceCube and are well understood: the measurement of the atmospheric neutrino spectrum with IceCube in its 40-string configuration is discussed in~\citet{Abbasi:2010ie}, the atmospheric muon energy spectrum measured with the 22-string configuration is presented in~\citet{Berghaus:2009zk}.
The flux of atmospheric muons exceeds the flux of atmospheric neutrinos by 5 orders of magnitude. 
The background of atmospheric muons can be reduced significantly by restricting the neutrino search to the northern hemisphere. 
Muons from the northern hemisphere cannot penetrate the Earth and reach the detector. However, a small fraction of the southern hemisphere muons are mis-reconstructed, i.e. truly down-going (entering the detector from above) but reconstructed as up-going. Owing to the large flux of atmospheric muons, these mis-reconstructed muons represent a significant contamination. Imposing requirements on the event reconstruction quality allows a suppression of the mis-reconstructed muon background.
%\end{linenumbers}

\subsection{Neutrino Selection Criteria and Efficiency}
\label{subsection:eventSelection}
\begin{figure}[t]
\centering
  \includegraphics[angle=0,width=8.5cm, trim=0cm 0cm 0cm 0cm, clip=true]{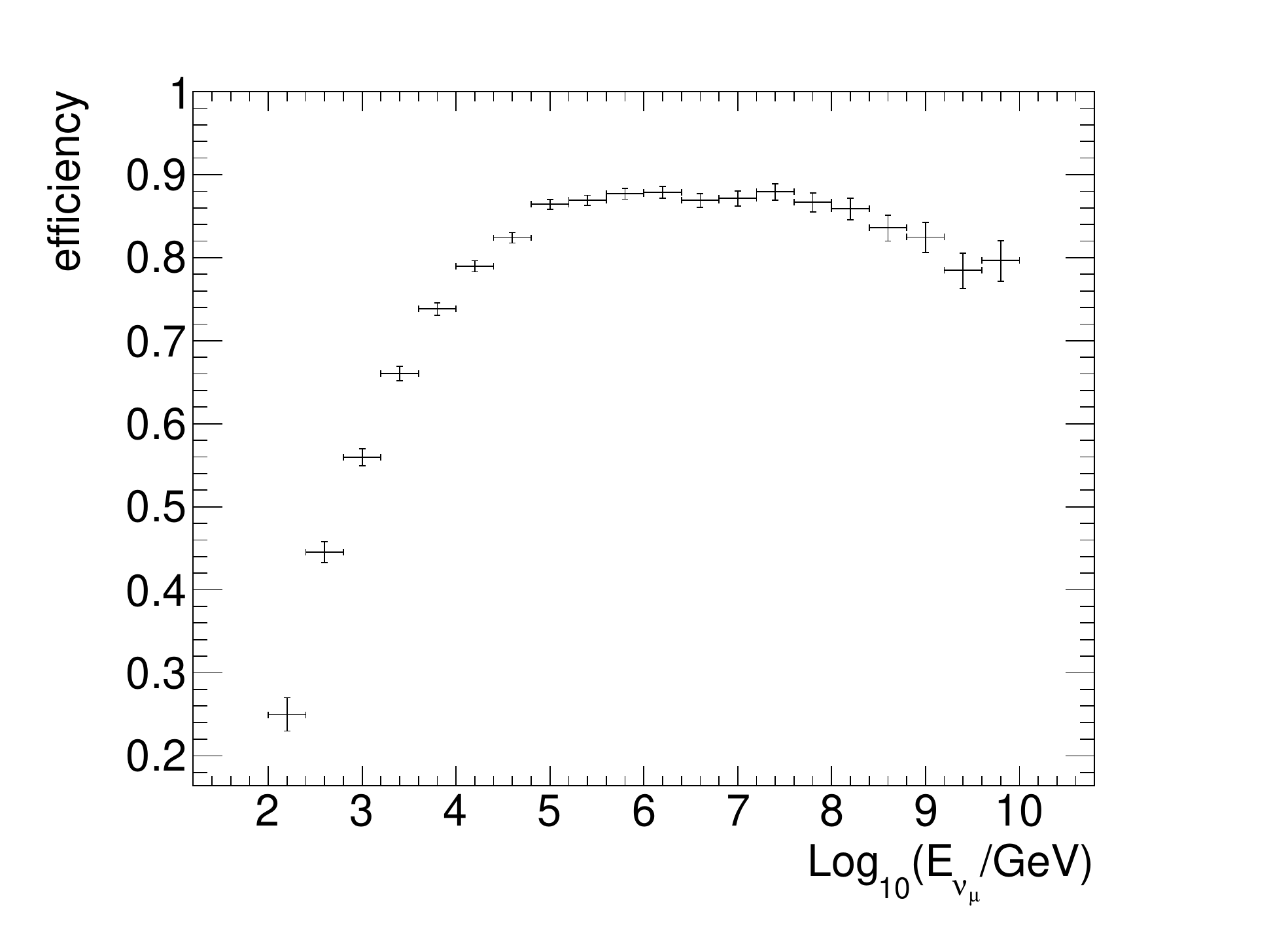}
  \caption{Filter efficiency of Level 3 relative to Level 2 as function of energy for well-reconstructed events (IC59).}
  \label{pic:effVsE}
\end{figure}

\begin{figure}[t]
\centering
  \includegraphics[angle=90,width=8.5cm, trim=0cm 0cm 0cm 0cm, clip=true]{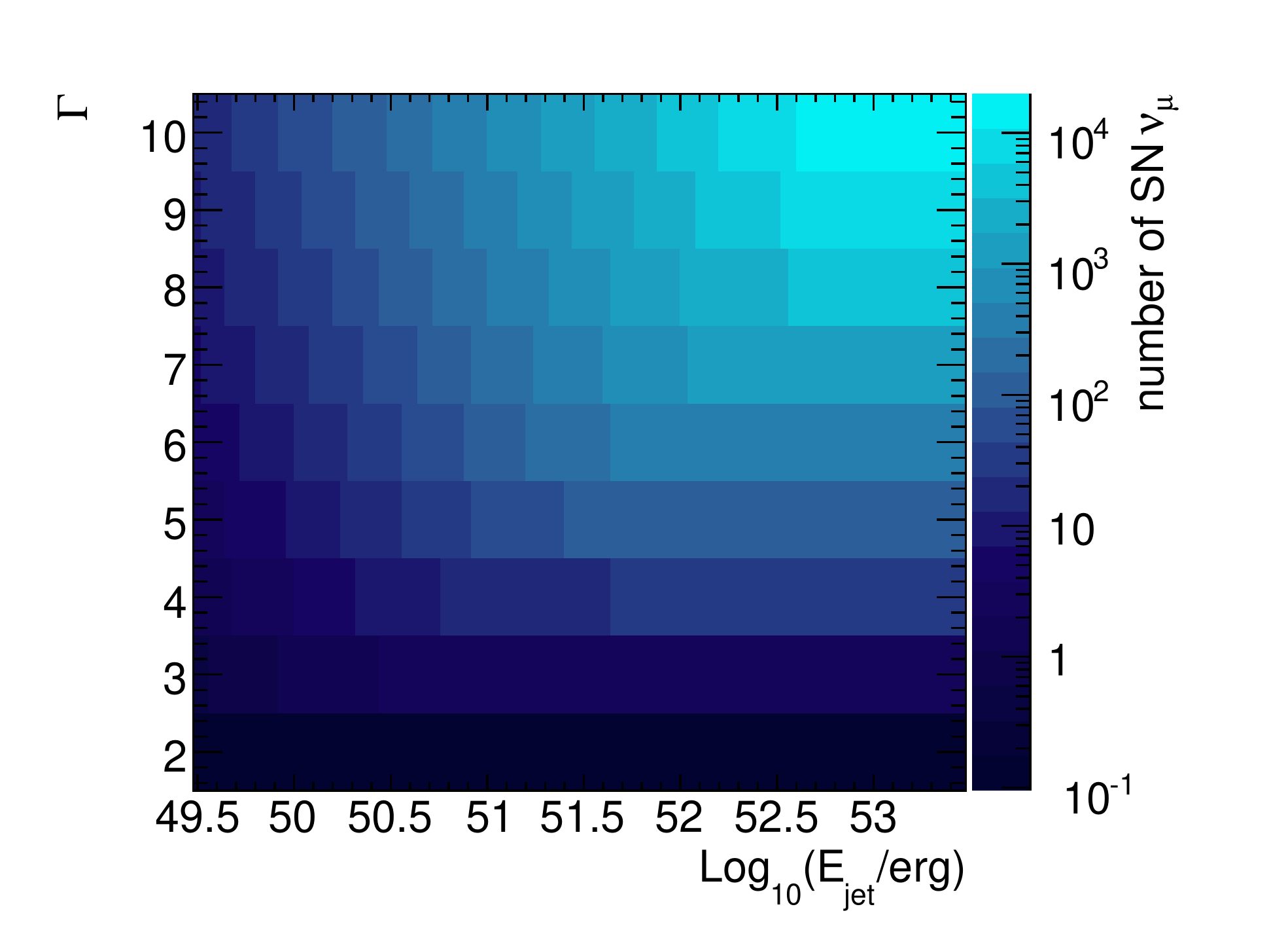}
  \caption{Expected number of SN neutrinos from a SN at distance $10$\,Mpc with a jet pointing towards Earth for different model parameter combinations (IC59).}
  \label{pic:ABEventsPerSN}
\end{figure}

\begin{figure}[t]
\centering
  \includegraphics[angle=0,width=8.5cm, trim=0cm 0cm 0cm 0cm, clip=true]{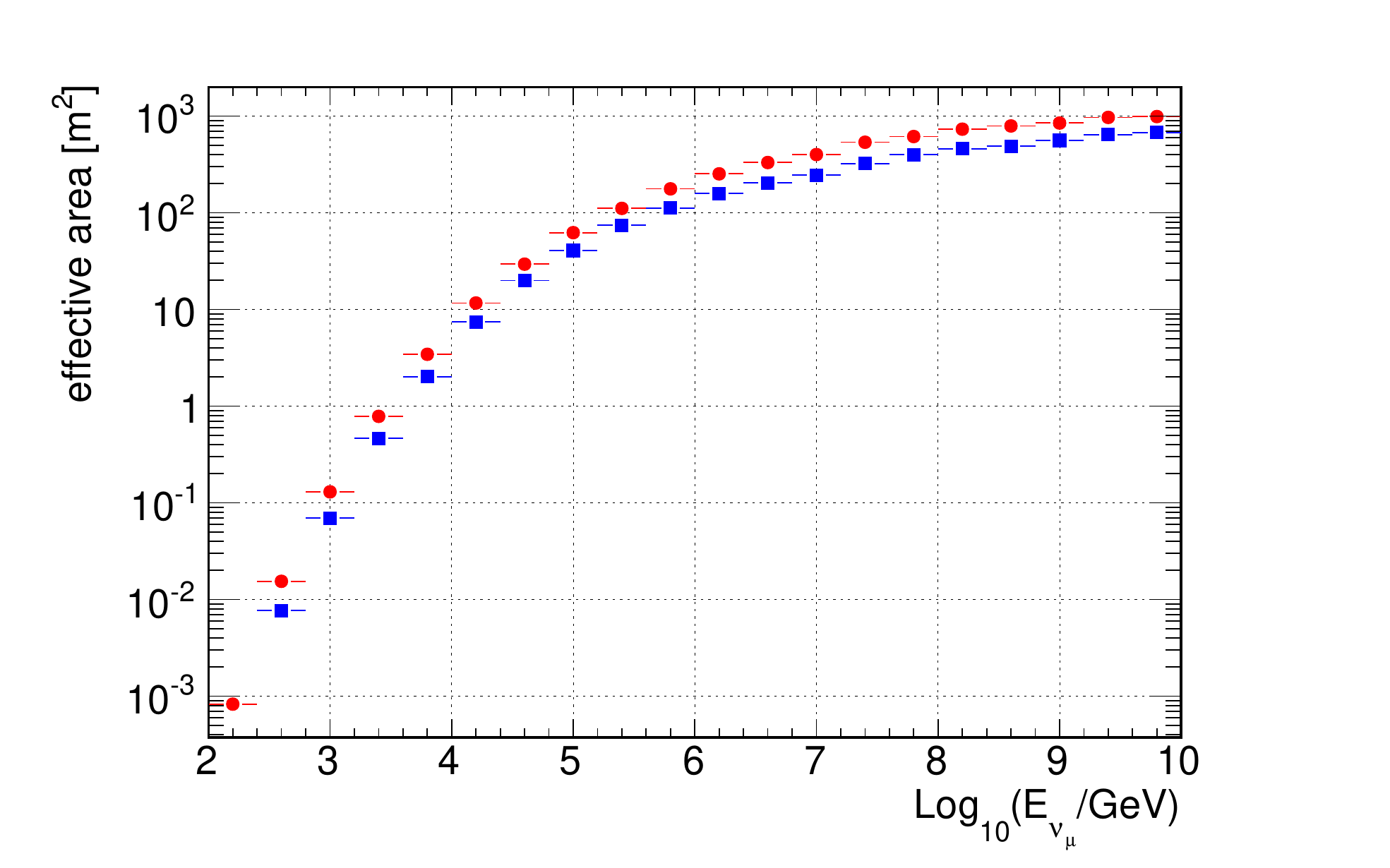}
  \caption{Neutrino effective area at the final selection level for well-reconstructed events for IC40 (blue squares) and IC59 (red circles).}
  \label{pic:effArea}
\end{figure}

\begin{figure}[t]
 \centering
 \includegraphics[angle=0,width=8cm]{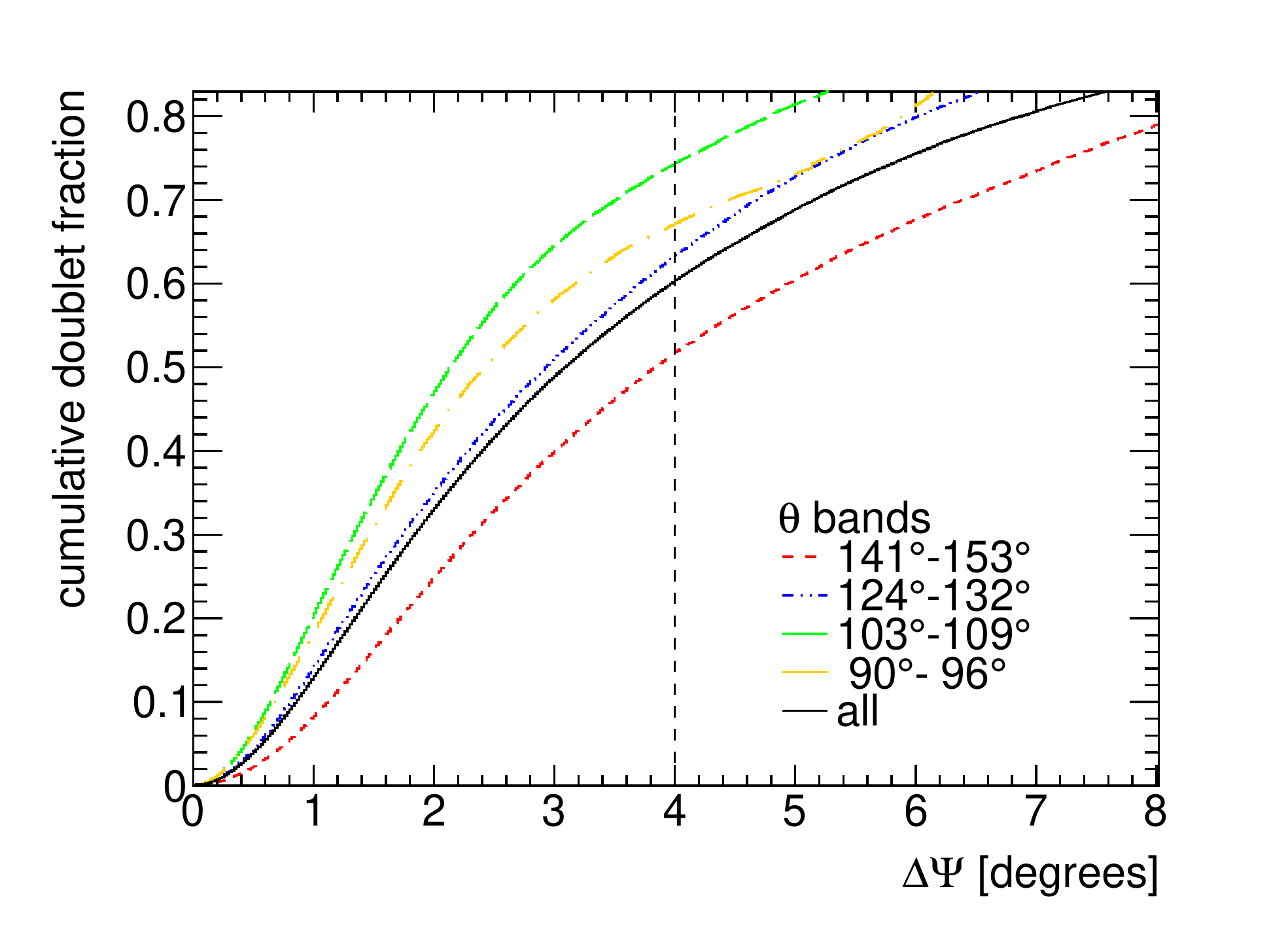}
  \caption{Angular difference between the reconstructed directions of two neutrinos with identical true direction for jet parameters $\Gamma=3$ and $E_{\rm{jet}}=3\times10^{51}$\,erg (IC59).}
  \label{pic:doubletAngWindow}
\end{figure}

%\begin{linenumbers}
The expected neutrino signal according to the soft jet SN model can be calculated as a function of two model parameters: the Lorentz boost factor $\Gamma$ and the jet energy $E_{\rm{jet}}$ (see section~\ref{sec:neutrinoFlux}).
Signal events are simulated following the predicted neutrino flux spectrum in order to develop and optimize selection criteria to distinguish signal and background events.
To suppress the background of atmospheric neutrinos, which we cannot distinguish from the soft
SN neutrino spectrum, we require the detection of at least two events within $\Delta t = 100$\,s and an angular
difference between their two reconstructed directions of $\Delta\Psi \leq 4^{\circ}$. The choice of the size of the time window
is motivated by the duration of the jet, i.e. the activity of the central engine, which is typically $10$\,s~\citep{Razzaque:2005bh}.
%by the jet penetration time. 
The observed gamma-ray
emission from long GRBs has a typical length of
$50$\,s~\citep{Gehrels:2009qy}, which roughly corresponds to the time for %duration of
a highly relativistic jet to penetrate the stellar envelope.
The angular window $\Delta\Psi$ is determined by the %IceCube's
angular resolution of IceCube and was optimized along with the other selection parameters.
The final set of selection cuts has been optimized in order to reach a background multiplet rate of \til$25$ per year corresponding to the maximal number of alerts accepted by ROTSE. 
Combining the neutrino measurement with the optical measurement allows the cuts to be relaxed compared to the neutrino point source analysis with IceCube \citep{PointSourceIC40} yielding a larger background contamination but at the same time a higher signal passing rate.\\
One doublet is not significant by itself, but may become significant when the optical information is added. 
%Each multiplet is forwarded to the
%telescopes. %, which grant an observing time corresponding to $\sim30$ alerts per year. \\
From the maximal allowed background multiplet rate %allows us to estimate
a corresponding maximal singlet rate $R_{1}$ can be estimated as follows. 
The probability to obtain a background triplet (three atmospheric neutrinos arriving by accident within 100\,s and within $\Delta \Psi$) or any multiplet of higher order is negligible, we therefore only consider doublets. Requiring no more than 25 background doublets per year ($R_{2}\lesssim25$\,year$^{-1}$) corresponds to a rate of isotropic background singlets of:
%\end{linenumbers}
% \begin{eqnarray}
%  R_{1} &\lesssim& \sqrt{R_{2} \frac{\Omega_{\rm{north}}}{\Delta t \cdot \Delta \Omega} } \nonumber \\
%        &=&\sqrt{25\rm{\,year}^{-1} \frac{20627(^{\circ})^2}{100\textrm{\,s} \cdot (\Delta\Psi)^2\cdot \pi} }=\frac{7.2\rm{\,mHz}}{\Delta\Psi},
% \end{eqnarray}
\begin{equation}
 R_{1} \lesssim  \sqrt{R_{2} \frac{\Omega_{\rm{north}}}{\Delta t \cdot \Delta \Omega} } = \frac{7.2\rm{\,mHz}}{\Delta\Psi},
\end{equation}
%\begin{linenumbers}
where $\Delta \Omega = \pi (\Delta\Psi)^2$ is the solid angle defined by the doublet condition and $\Omega_{\rm{North}} = 20627(^{\circ})^2$ is the solid angle of the Northern sky, i.e. $2\pi$.
The event selection is optimized in order to restrict the singlet rate to $7.2\rm{\,mHz}/\Delta\Psi$ while obtaining the best signal passing rate. $\Delta\Psi = 4^{\circ}$ was found to be the best choice during the cut optimization. This corresponds to a singlet rate of 1.8\,mHz.
The event selection is based on one of the standard IceCube muon event filters (in the following referred to as Level~1), which is commonly used by several offline analysis. It is discussed in detail in \citet{PointSourceIC40}. The Level~1 filter selects muon tracks, which 
are reconstructed as up-going (passing through the Earth) based on fast and simple algorithms. It selects \til$2$\% of all triggers with a signal efficiency of $90\%$ for up-going neutrinos following an $E^{-2}$-spectrum. It is still largely dominated by atmospheric muons. To the Level~1 filter stream we apply additional selection criteria
mainly based on the reconstruction quality. This yields the so-called Level~2 filter stream.
The significantly smaller rate of \til$3$\,Hz allows us to perform time consuming reconstructions online, which 
provide a more accurate estimate of the event direction.
These reconstructions are based on a muon-likelihood function described in \citet{AMANDAReco}, which parametrizes the probability of observing
the spatial distribution and timing of the hits with respect to a muon track hypothesis. 
The negative logarithm of the likelihood, $-\log \mathcal{L}$, is minimized, i.e. the likelihood, $\mathcal{L}$, is maximized by varying the track direction to yield the best-fit direction and position
for the muon track. Iterative fits repeat the minimization with a different initial track hypothesis to reduce the problem of local minima. The iteration with the smallest minimum is the final fit result. At Level~2, a ten-fold iterative likelihood reconstruction is available. % (10 iterations) and a paraboloid reconstruction.
%The latter fits a paraboloid to the likelihood landscape around the minimum defined by the best fit
%parameters yielding an estimate of the reconstruction accuracy, $\sigma$. 
%It will be referred to as $\sigma$. 
The uncertainty on the reconstructed direction, $\sigma$, is obtained from a fit of a paraboloid to the $-\log \mathcal{L}$ space around the direction. %, where $\mathcal{L}$ is the likelihood obtained by the reconstruction algorithm. 
Based on these more sophisticated reconstruction algorithms we select
our final event stream, Level 3: a powerful parameter to reject
mis-reconstructed events is given by the reduced negative logarithm of the likelihood, 
$-\log \mathcal{L} / (\rm{N}_{\rm{Ch}}-5)$ or a modified version given by
$-\log \mathcal{L} / (\rm{N}_{\rm{Ch}}-2)$ or $-\log \mathcal{L} / (\rm{N}_{\rm{Ch}}-3.5)$, where N$_{\rm{Ch}}$ is the number of triggered DOMs after time window cleaning.\\
%combination of the number of triggered DOMs, N$_{\rm{Ch}}$, and the negative logarithm of the
%likelihood obtained by the reconstruction algorithm, $-\log(\mathcal{L})$. 
%In IC40 $\log(\mathcal{L})/(\textrm{N}_{\textrm{Ch}}-5)$
%was used while for IC59 $\log(\mathcal{L})/(\textrm{N}_{\textrm{Ch}}-3.5)$ showed a better selection power.
A large number of hits with a small time residual, N$_{\rm{Dir}}$, i.e.\ registered within a time window 
[-25\,ns, 75\,ns] relative to the expected arrival time for unscattered light given by the track geometry,
ensures a good track reconstruction quality, since photons causing 
those direct hits are less affected by scattering. % and a large number of direct hits therefore ensures a good track reconstruction quality.
Furthermore, %the length of the projection of the direct hits on the track L$_{\rm{Dir}}$ indicates a high track quality.
the maximum distance, L$_{\rm{Dir}}$, along the reconstructed
track direction between any two hits with small time residual ([-25\,ns, 75\,ns]) is a measure of the track quality.
%Furthermore, the number of direct hits, N$_{\rm{Dir}}$ and the length of their projection to the track, 
%L$_{\rm{Dir}}$, are used. As direct hits we define those hits, which were registered within a time window 
%[-25\,ns, 75\,ns] relative to the expected arrival time given by the track geometry. Photons causing 
%direct hits are less effected by scattering and a large number of direct hits therefore ensures
%a good track reconstruction quality. 
In addition, only up-going events with zenith angle $\theta \geq 85^{\circ}$ (IC40) and $\theta \geq 90^{\circ}$ (IC59) are selected\footnote{Southern hemisphere events have zenith angles of $0^{\circ} < \theta < 90^{\circ}$ in the coordinate system of IceCube.}. In IC40, an additional cut on the doublet direction was applied selecting only doublets with a combined zenith of $\theta_{\rm{Doublet}} \geq 90^{\circ}$.
A single-iteration likelihood reconstruction (llh1) and a second two-iteration
likelihood reconstruction (llh2) using the seed track and the inverted seed track of llh1 were used in IC40. The ten-fold iterative likelihood
reconstruction (10it) was used in both IC40 and IC59.\\
The final cuts at Level~3 for IC40 are
%\end{linenumbers}
\begin{center}
$\theta_{\rm{llh1}}\geq85^{\circ} \textrm{ AND }
\theta_{\rm{llh2}}\geq85^{\circ} \textrm{ AND }
\theta_{\rm{10it}}\geq85^{\circ} \textrm{ AND }$ \\
$-\log{\mathcal{L}}/(\rm{N}_{\rm{Ch}}-5)\leq8.85 \textrm{ AND }$ \\
$((\rm{N}_{\rm{Dir}}\geq7 \textrm{ AND } \rm{L}_{\rm{Dir}}\geq225) \textrm{ OR } \rm{N}_{\rm{Ch}}\geq200)$
\end{center}
%\begin{linenumbers}
and for IC59
%\end{linenumbers}
\begin{center}
$\theta_{\rm{10it}}\geq90^{\circ} \textrm{ AND }
-\log{\mathcal{L}}/(\rm{N}_{\rm{Ch}}-3.5) \leq 7.7 \textrm{ AND }$\\
$((\rm{N}_{\rm{Dir}}\geq7 \textrm{ AND } \rm{L}_{\rm{Dir}}\geq250) \textrm{ OR } \rm{N}_{\rm{Ch}}\geq100)$.
\end{center}
%\begin{linenumbers}
The spectrum obtained from the AB05 model 
can be either hard or soft depending on the choice of model parameters. The choice of cuts specified above is a compromise yielding adequate passing rates for
all considered model parameters. 
%Therefore the cuts are 
%optimized for both, a hard $E^{-2}$ and a soft $E^{-3}$ spectrum 
%resulting in two different sets of cuts. A compromise between the two sets of cuts has been chosen 
%in order to provide good passing
%rates for all choices of model parameters. 
Figure~\ref{pic:effVsE} shows the energy dependence of the Level 3 filter efficiency relative to Level 2 for well-reconstructed events ($|\overrightarrow{\Psi}_{\rm{true}}-\overrightarrow{\Psi}_{\rm{reco}}|<3^{\circ}$ and $\theta_{\rm{true}}>90^{\circ}$, where the unit vector $\overrightarrow{\Psi}$ indicates the track direction). The filter efficiency is defined as the fraction of simulated signal events passing the filter. For energies above $100$\,TeV the filter is $90$\% efficient while for lower energies the efficiency decreases to $50$\% at $1$\,TeV and $20$\% at $100$\,GeV.
The cuts were adjusted from IC40 to IC59 to account for the larger detector volume in order to obtain a similar data passing rate. 
Table~\ref{tab:cuts} shows the data passing rates for IC40 and IC59 at different cut levels. Furthermore, the table contains the expected number of detected SN neutrinos for a SN at distance $d_{\rm{SN}}=10$\,Mpc with a jet of energy $E_{\rm{jet}}=3\times10^{51}$\,erg, which points towards Earth, and two choices of the boost factor $\Gamma$ ($4$ and $10$). The expected number of well-reconstructed SN neutrinos is given in brackets since only these events are useful to trigger optical telescopes. Figure~\ref{pic:ABEventsPerSN} shows the expected number of events for different model parameter configurations for IC59.
%To better compare IC40 and IC59, we applied an additional zenith cut $\theta_{\rm{10it}}\geq90$ to the IC40 data. Note that the real cut was not applied to the single events zenith angle, but to the doublet direction zenith angle.
%\end{linenumbers}
\begin{table*}[t]
 \caption{\sf{Cut Levels and Event Rates. Values in brackets refer to well-reconstructed events.}}
  \label{tab:cuts}
 \begin{center}
  \begin{tabular}{ l l l c c c c}
     \toprule
    \textbf{Cut Level}  & \multicolumn{2}{c}{\textbf{Event Rate}}  & \multicolumn{4}{c}{\textbf{Neutrino events for SN at $d_{\rm{SN}}=10$\,Mpc}}  \\
                        &            &                    & \multicolumn{2}{c}{$\Gamma=4$, $E_{\rm{jet}}=3\times10^{51}$\,erg}   & \multicolumn{2}{c}{$\Gamma=10$, $E_{\rm{jet}}=3\times10^{51}$\,erg} \\
                   & IC40  & IC59   & IC40 & IC59  & IC40 & IC59 \\
    \midrule
    Level 1       & 20.7\,Hz &  22.7\,Hz      &  68.0 (18.6) & 133.3 (34.7)   & 3385.5 (1081.3) & 5304.0 (1877.7) \\
    Level 2       & 2.74\,Hz &  3.32\,Hz      &  48.1 (17.6) & 100.0 (32.9)   & 2544.6 (1076.0) & 4225.9 (1801.1) \\
    Level 3       & 2.17\,mHz & 1.86\,mHz     &  13.3 (8.7) & 22.4 (16.3)    & 947.4 (674.6) & 1441.7 (1153.0) \\
    \bottomrule
  \end{tabular}
 \end{center}
\end{table*}
%\begin{linenumbers}
Table~\ref{tab:cuts} shows that we expect many more neutrino events for a SN with a high boost factor. On the other hand, a large boost factor implies a small jet opening angle ($\theta \propto 1/\Gamma$) and hence a smaller probability of the jet pointing towards Earth. Furthermore, one sees an increase in the expected number of SN neutrinos at Level 3 from IC40 to IC59. The detector volume increased by roughly $50\%$. However, owing to improved performance of the reconstruction algorithms applied to data of the larger IC59 detector, the signal passing rate at Level 3 increased by 52\% ($\Gamma=10$) and 68\% ($\Gamma=4$). The increase for well-reconstructed events is even higher (71\% for $\Gamma=10$ and $87$\% for $\Gamma=4$). %, showing that we are very efficient in selecting well-reconstructed events with the larger detector.
% This can be explained by harder cuts which need to be applied to the IC59 data set in order to reach a similar final background doublet rate. 
The neutrino effective area for well-reconstructed events at final cut level is shown in Fig.~\ref{pic:effArea} for IC40 and IC59.
%\begin{linenumbers}
The Level 3 cuts reduce the singlet rate to \til$1.8$\,mHz for IC59. The IC40 rate is slightly higher at $2.2$\,mHz, but an additional cut on the doublet direction ($\theta_{\rm{doublet}}\leq90^{\circ}$) ensures a multiplet rate of less than 25 per year. 
The Level 3 data stream consists of $37$\% ($70$\%) atmospheric neutrinos for IC40 (IC59), the rest is a contamination of mis-reconstructed atmospheric muons.
%\end{linenumbers}
% \begin{figure}[b]
% \centering
%   \includegraphics[angle=0,width=8.5cm, trim=0cm 0cm 0cm 0cm, clip=true]{images/efficiencyVsEnergy_RelToL2_WR_IC59.pdf}
%   \caption{Filter efficiency of Level 3 relative to Level 2 as function of energy for well-reconstructed events (IC59).}
%   \label{pic:effVsE}
% \end{figure}
% \begin{figure}[b]
% \centering
%   \includegraphics[angle=0,width=8.5cm, trim=0cm 0cm 0cm 0cm, clip=true]{images/ABEventsIC59.pdf}
%   \caption{Expected number of SN neutrinos from SN at distance $10$\,Mpc with a jet pointing at Earth for different model parameter combinations (IC59).}
%   \label{pic:ABEventsPerSN}
% \end{figure}
% \begin{figure}[b]
% \centering
%   \includegraphics[angle=0,width=8.5cm, trim=0cm 0cm 0cm 0cm, clip=true]{images/effAreaIC40andIC59WRNew.pdf}
%   \caption{Effective area for well-reconstructed events for IC40 (blue) and IC59 (red).}
%   \label{pic:effArea}
% \end{figure}
% \begin{linenumbers}
In addition -- as described above -- we require at least two events to arrive within $\Delta t=100$\,s and with an angular distance of $\Delta\Psi \leq 4^{\circ}$ to reduce the background of atmospheric neutrinos.
The signal efficiency of the angular coincidence cut for different zenith bands is displayed in 
Fig.~\ref{pic:doubletAngWindow}.
\begin{figure}[t]
\centering
\includegraphics[angle=0,width=8cm]{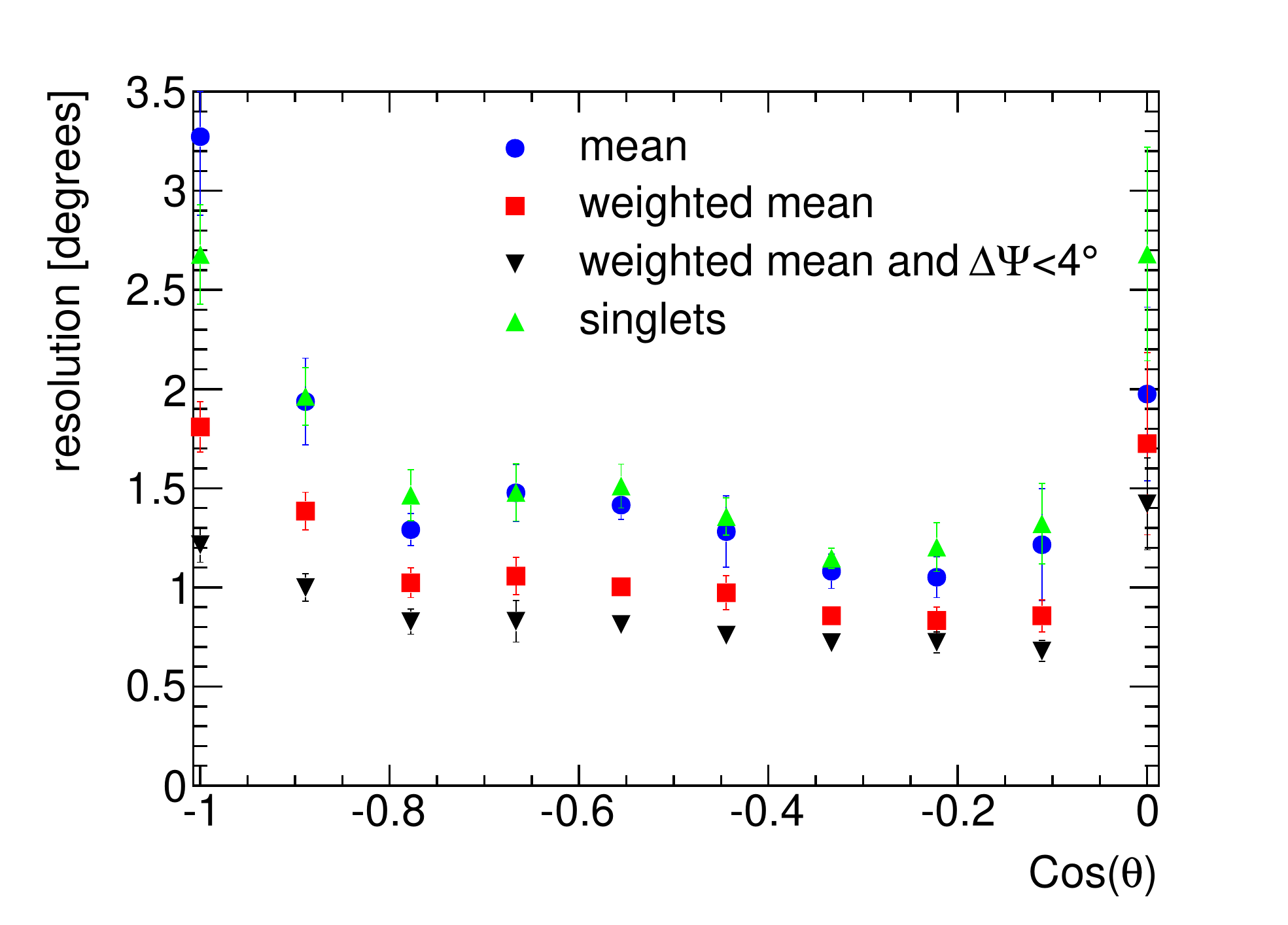}
  \caption{Doublet resolution using an ordinary mean (blue circles), a weighted mean (red squares) compared to the singlet resolution (green triangles) for signal neutrinos ($\Gamma=3$, $E_{\rm{jet}}=3\times10^{51}$\,erg). Applying the directional coincidence cut $\Delta \Psi < 4^{\circ}$ (black triangles) keeps mainly well-reconstructed doublets and yields a further improvement of the doublet resolution (IC59).}
  \label{pic:doubletResZen}  
\end{figure}
\begin{figure} 
\centering
   \includegraphics[angle=0,width=8cm]{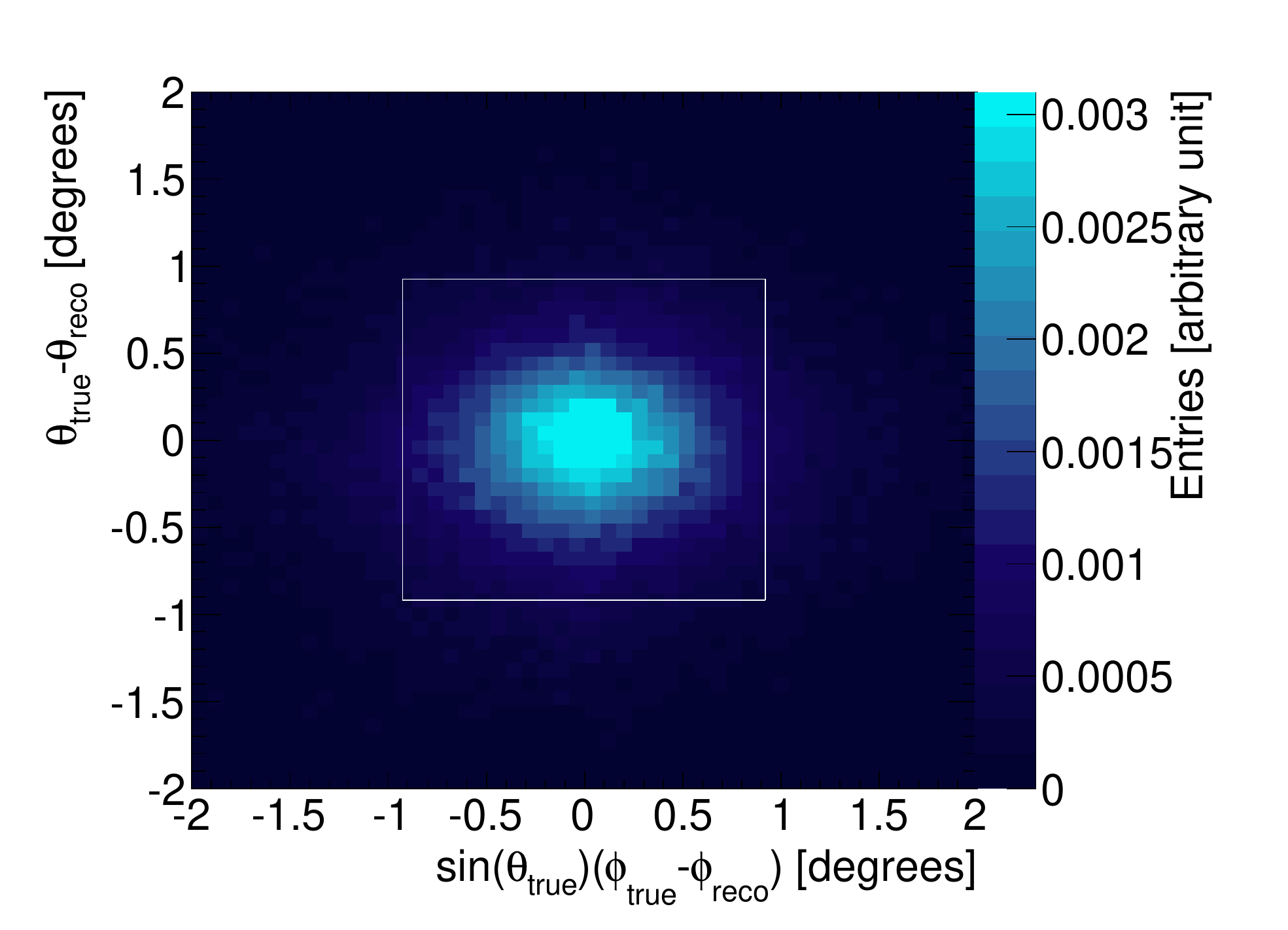}
  \caption{Deviation of reconstructed multiplet direction from true direction (IC59). The black box shows ROTSE's
 	   field of view of $1.85^{\circ} \times 1.85^{\circ}$.}
  \label{pic:ROTSEFoV}
\end{figure}
%The ROTSE telescopes have a FoV of $1.85^{\circ} \times 1.85^{\circ}$. 
The average passing rates depend on the assumed energy spectrum (i.e.\ the model parameters) and are in the range of $56-69$\% (IC40) and $60-74$\% (IC59) for $1\leq \Gamma \leq 10$ and $3.1\times10^{49}$\,erg $ \leq E_{\rm{jet}} \leq 3\times10^{53}$\,erg. The reconstruction accuracy has improved because of the increased detector volume. Large reconstruction uncertainties might lead to mis-pointing of the telescopes and in the worst case the real source position might lie outside ROTSE's field of view of $1.85^{\circ} \times 1.85^{\circ}$. To improve the accuracy of the direction forwarded to the telescopes, the doublet direction is calculated as a weighted mean from the directions of the individual events in the multiplet.
%from the single reconstructed directions comprising the multiplet. 
The single events are weighted with $1/\sigma^2$, where $\sigma$ is the reconstruction error estimated by the paraboloid fit. %, i.e. well reconstructed events contribute stronger to the final direction. 
Compared to single events, doublets have a better resolution. The weighting improves the doublet resolution as can be seen in Fig.~\ref{pic:doubletResZen}. It further improves after applying the directional coincidence condition $\Delta \Psi \leq 4^{\circ}$. Figure~\ref{pic:ROTSEFoV} shows the doublet point spread function in $\theta$-$\phi$-space compared to ROTSE's FoV. Depending on the model parameters, 41-53\% (IC40) and 44-61\% (IC59) of all doublet events lie within ROTSE's field of view, providing a good match for this search.
%\end{linenumbers}

\section{Search for Optical Counterparts}
\label{sec:OpticalCounterpart}
%\begin{linenumbers}
The IceCube multiplet alerts are forwarded to
the robotic optical transient search experiment (ROTSE), which consists of four 
identical telescopes located in Australia, Texas, Namibia and Turkey~\citep{ROTSE}. 
The telescopes stand out because of their large FoV of $1.85^{\circ} \times 1.85^{\circ}$ 
and a rapid response with a typical telescope slew time of 4\,s to move the telescope 
from the standby to the desired position.
The telescopes have a parabolic primary mirror with a diameter of 45\,cm. 
To be sensitive to weak sources no bandwidth filter is used. 
ROTSE is most sensitive in the R-band (\til$650$\,nm). The wide field of view is imaged onto a back-illuminated thinned CCD with $2048\times2048$ $13.5$\,$\mu$m pixels. %Note that ROTSE's FoV matches well 
%the point spread function of IceCube as shown in figure~\ref{pic:ROTSEFoV}.
The camera has a fast readout cycle of $6$\,s and 
is cooled to $-40^{\circ}$\,C in order to reduce thermal noise. For a $60$\,s 
exposure at optimal conditions the limiting magnitude is around $m_{R}\approx 18.5$, 
which is well suited for a study of GRB afterglows during the first hour (up to one day for very bright afterglows) %or more 
and SN light curves with apparent peak magnitude $\leq16$. 
\begin{figure}[t]
 %\centering
  \includegraphics[angle=0,width=9cm, trim=0cm 0cm 0cm 0cm, clip=true]{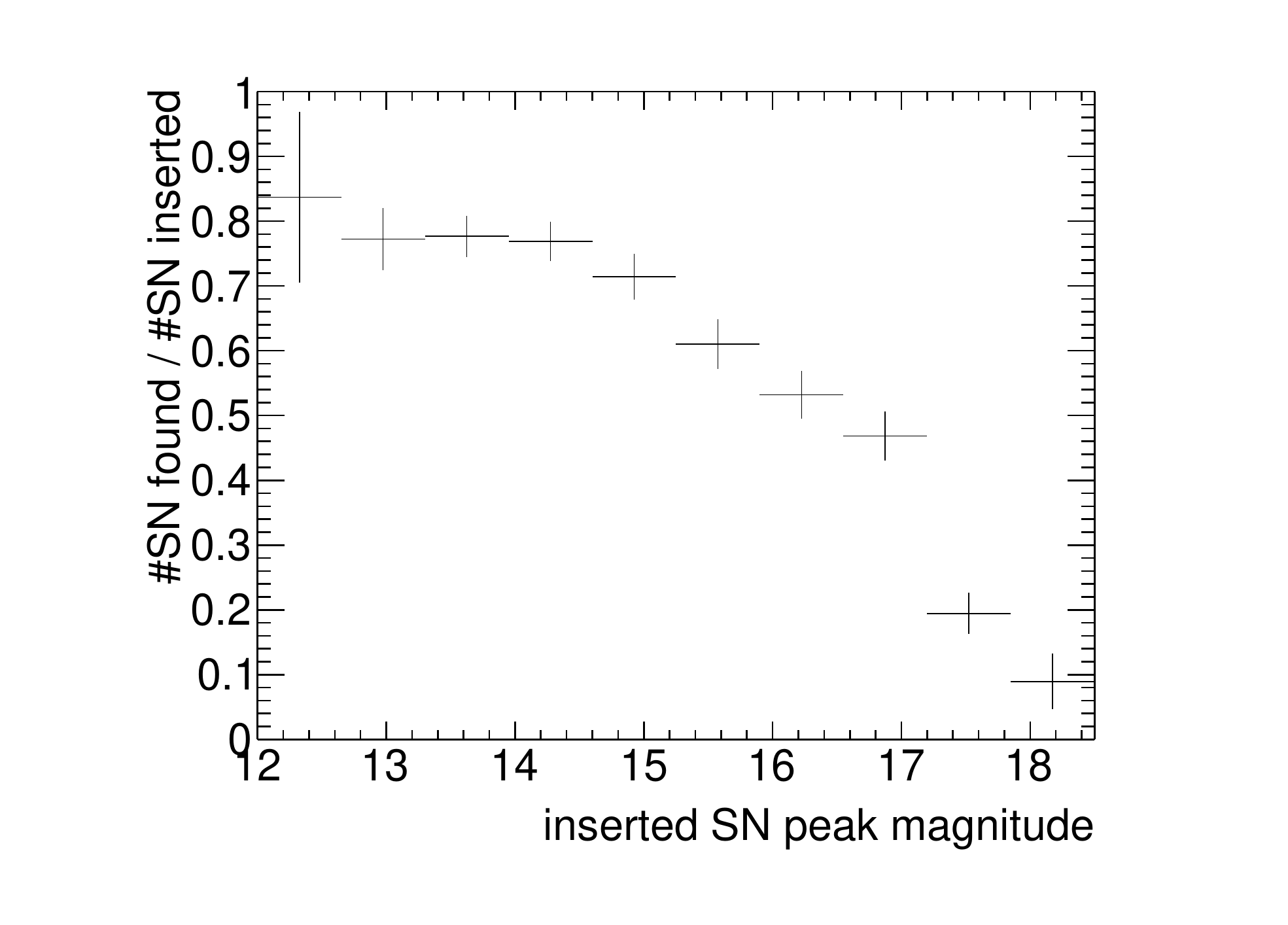}   
  \caption{Efficiency to find inserted SNe as a function of the apparent peak magnitude averaged over all alert directions assuming a template light curve.}
  \label{pic:ROTSEEffMag}
\end{figure}
The corresponding FWHM (full width at half maximum) of the stellar images is 
$<2.5$~pixels ($8.1$ arcseconds). The telescopes are operated robotically and managed by a fully-automated system. 
%of interacting software daemons within a Linux environment. 
Observations are scheduled in a queue 
and are processed in the order of their assigned priority.  
IceCube triggers have second highest priority
after GRB follow-ups triggered by the GRB Coordinate Network (GCN). \\
%IceCube trigger are sent to the telescope computer via a socket connection from 
%a dedicated server in at the University of Wisconsin. \\
Once an IceCube alert is received by one of the telescopes,
a predefined observation
program is started and the corresponding region of the night sky will
be observed within seconds: the prompt observation includes
thirty exposures of 60 seconds length, follow-up
observations are performed for 14 nights. This was extended on Oct. 27, 2009 to 24 nights, with daily observations for 12 nights and then observations during every second night up to day 24 after the trigger was received. Eight images
with 60 seconds exposure time are taken per night. The
prompt observation is motivated by the typical rapidly decaying
light curve of a GRB afterglow, while the follow-up
observations during 14 or 24 nights permit the identification of
a rising SN light curve. With IC40 and IC59, the online processing latency of several hours made the search for an optical GRB afterglow unfeasible. We therefore focus on the SN light curve detection in the ROTSE data.\\
Image correction and calibration are performed at the telescope sites. 
The images of each night are co-added in order to obtain a deeper image. Co-adding includes a geometrical transformation to correct for optical distortions and mis-alignment of the images before the pixel contents are added. A reference image is subtracted from each co-added image.
As deep images are usually not available for the positions we would like to observe, we 
initially choose the deepest image of our observing sequence as the reference image (due to different weather conditions in different nights the image quality varies). The source could be present in the reference image, but it would be of different brightness compared to earlier or later images, and both positive and negative deviating sources in the subtracted images will be detected. 
If no early image of good quality was available to be selected as reference image (30\% of the alerts) we take another deep image several months later. 
%In 70\% of the alerts we took another deep image roughly one year later (for the
%remaining 30\% the position was not observable by ROTSE).
Both SN light curves and GRB afterglows would have faded after a few weeks and would not 
be present in the newly taken reference image. \\
The applied subtraction algorithm was developed by \citet{CrossConv}. Both the new and the reference image are folded by a kernel function in order to match their point spread functions (PSFs). The convoluted images then allow a pixel by pixel 
subtraction. The software \texttt{SExtractor} (source extractor), \citet{sextractor}, extracts all objects from an image by first determining the background and then identifying clusters of pixels with a significance $>1\sigma$ above the background level. All extracted objects found in the subtracted images with signal-to-noise ratio larger than $5$ are candidates for variable sources. %, but are also frequently caused by mis-subtractions.
However, bad image quality, failed image convolution, bad pixels and other effects frequently cause artifacts in the subtraction process, requiring further selection of the candidates.
A candidate identification algorithm including a boosted decision tree is applied to classify candidates according to geometrical and variability criteria. The algorithm was trained using a signal of 
%simulation. Our expected optical signal is 
a SN light curve starting at the time of neutrino
detection. We use a SNIbc template light curve based on SN1999ex \citep{SN1999ex}. The time of the shock
break-out was measured for SN1999ex and provides a time stamp for the explosion, i.e.\ the start of the light curve, which can be associated with high-energy neutrino emission. To simulate the SN, fake stars are inserted on top of galaxies in every single image from the observation sequence with a brightness according to the SN light curve template. 
To ensure that the PSF of the inserted star reflects all the features of the PSF of existing stars in the image we use the mean PSF of all stars in a $291\times291$ pixel box around the insertion coordinates. The PSF of a single star is obtained from a box of $15\times15$ pixel around the star's center.
The manipulated images are processed with the same pipeline as described above.
The signal sample consisting of inserted fake SNe is used to train the classification algorithm as well as to estimate ROTSE's efficiency.
The efficiency is given by the fraction of inserted SNe, that has been detected by the processing and candidate identification.
For some inserted SNe the detection algorithm fails: If the quality of the image is bad (e.g.\ large average
FWHM or small limiting magnitude) the image convolution performs badly. Candidates close to 
saturated objects or close to objects listed in the two micron all sky survey (2MASS) point source catalog, which are very likely stars, are removed automatically.
Figure~\ref{pic:ROTSEEffMag} shows the fraction of simulated SNe that are found by the algorithm as a function of the inserted apparent peak magnitude.
The efficiency as a function of the apparent peak magnitude can be converted to the efficiency as a function
of SN distance $\varepsilon_{\rm{ROTSE}}(d)$ (see Fig.~\ref{pic:ROTSEEffDist}) assuming an
absolute R-band magnitude of $M = -18\pm1$ for core-collapse SNe~\citep{SNMag}. The relation of distance and magnitude is given by
\begin{equation}
 m = M + 5 \left(\log_{10} \frac{d}{1pc} - 1\right),
\end{equation}
where $m$ is the relative magnitude.\\
%Out of 31 alerts
%forwarded to ROTSE 14 did not lead to an observation 
%(see section~\ref{sec:Results}).
%Therefore, a correction factor of 55\% has to be applied to obtain the final efficiency. 
The efficiency is used to calculate the number
of expected SNe detections for a given signal neutrino hypothesis. However, it
may also be used to estimate, for a given SN rate, the number of accidental
coincidences between a neutrino doublet and a SN detectable by ROTSE (see section~\ref{sec:Results}). 
%This efficiency is used to calculate the number of background SNe, i.e.\ the number of SNe, which are detected in coincidence 
%with a neutrino doublet by accident. 
%As not all SNe will be detected, it is used as well for calculating the number of expected SNe detections for a given signal neutrino prediction.
%Furthermore the efficiency becomes important for calculating the number of expected SN detections according to a model prediction. 
%not including the fraction of alerts without optical observations. 14 out of 31 alerts, which were forwarded to ROTSE, could not be observed because they were either too close to the sun, too close to the galactic plane or no good data could be collected due to bad weather or technical problems. E.g. a correction factor of 55\% has to be added to obtain the final efficiency.
%The efficiency as a function of the peak magnitude was converted to a function of SN distance
%$\varepsilon_{\rm{ROTSE}}(d)$ (see figure~\ref{pic:ROTSEEffDist}) assuming an absolute R-band
%magnitude of $-18\pm1$ for core-collapse SNe~\citep{SNMag}.
% \begin{figure}[t]
%  \centering
%   \includegraphics[angle=0,width=9cm]{images/FoundSNNew.pdf}   
%   \caption{Efficiency to find inserted SNe as a function of the peak magnitude.}
%   \label{pic:ROTSEEffMag}
% \end{figure}
\begin{figure}[t]
\centering
  \includegraphics[angle=0,width=9cm, trim=0cm 0cm 0cm 0cm, clip=true]{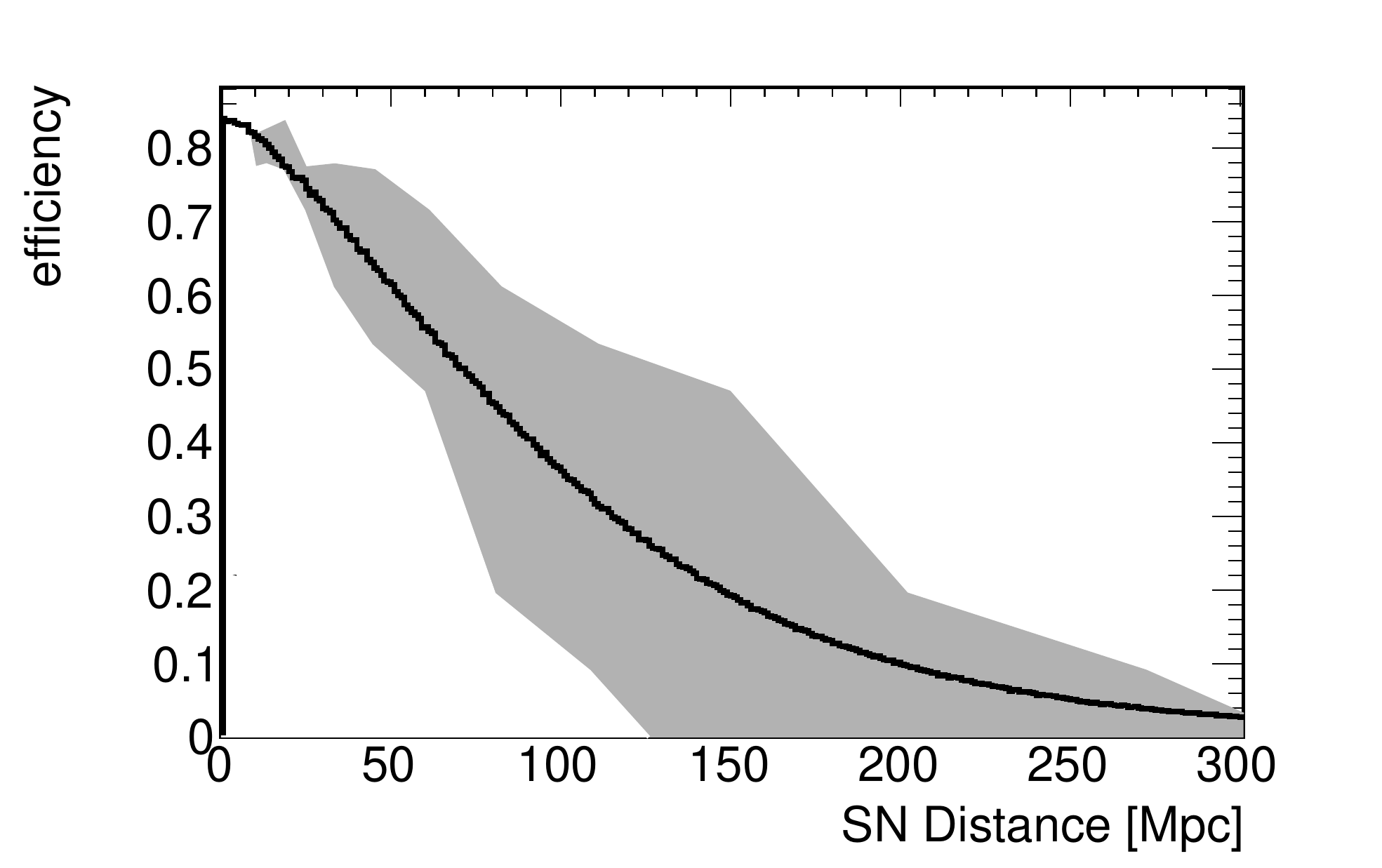}
  \caption{Black curve: Efficiency to detect core-collapse SN as a function of the distance to the SN assuming an absolute R-band magnitude of $-18\pm1$. Shaded region: lower bound assuming an absolute magnitude of $-17$, upper bound assuming $-19$. The breaks in the shaded regions are connected to the binning in Fig.~\ref{pic:ROTSEEffMag}. The binning effect is washed out in the black curve due to the assumed uncertainty in the absolute magnitude distribution of $\pm1$.}
  \label{pic:ROTSEEffDist}
\end{figure}
After application of the classification algorithm \til$10-200$ candidates for variable objects remain per alert depending on the quality of the images and the Galactic latitude. Fields close to the Galactic plane
contain a large number of stars, which complicates the subtraction and thus results in more candidates caused by subtraction artifacts. Tightening the cuts would 
reduce the number of candidates but at the same time reduce our sensitivity. 
The final candidates are summarized on a web page and are inspected visually by 
people trained on the signal simulation. 
%several trained (using the signal simulation) people. 
The web page shows a $100 \times 100$ (pixel)$^2$ extract of the new,
the reference and the subtracted image for each night. 
For comparison, an image from the digitized sky survey (DSS\footnote{\url{http://archive.stsci.edu/dss/}}) of the same patch of the sky is shown, which is deeper than the ROTSE images.
In addition, links to catalogs, such as NED\footnote{\url{http://ned.ipac.caltech.edu/}}, 2MASS \citep{2MASS} and SDSS \citep{SDSS} are provided. On the basis of this 
information the scanners have to decide whether the candidate is a SN, 
a variable star or a subtraction artifact. SN candidate identification by the human eye works well as shown in the galaxy zoo SN project \citep{GalaxyZoo}. 
%The visual scanning was performed by three
%individuals to ensure no good candidate was missed and to avoid false positives.
The visual scanning of the final candidates was carried out by three individual persons, who obtained similar results. All simulated SNe, which passed the computer selection, were identified in the scanning, i.e. the efficiency was $100\%$. Also the rate of false positives is expected to be small, because for a potential SN candidate its light curve and host galaxy properties would be inspected in detail. We hence assume, that the systematic uncertainty introduced by the scanning process is negligible. Note that in the future, for candidates identified in real time a spectrum can be obtained to ensure an unambiguous identification of the SN.
%No supernova signature was found in the optical data.
%\end{linenumbers}

\section{Systematic Uncertainties}
\label{sec:SysErrors}
%\begin{linenumbers}
Both the simulated neutrino sensitivity and the SN sensitivity are subject to systematic uncertainties.
%Systematic errors appear in the prediction of the signal neutrino flux as well in the prediction of SNe detections. Both will be discussed below. 
These systematic uncertainties are included in the limit calculation. In this limit calculation, Monte Carlo experiments are performed drawing the number of signal neutrino-events following a Poisson distribution (see appendix~\ref{appendixLimitCalculation} for details on the limit calculation). Systematic uncertainties are included by smearing the Poisson mean, i.e. the Poisson mean is multiplied by a factor following a Gaussian distribution with mean one and a width given by the systematic uncertainty. 
%Systematic uncertainties are included by drawing a random value for a specific uncertainty following a Gaussian distribution (the mean is zero and the width is given by the average uncertainty multiplied by the expected signal rate) and adding the resulting change in number of events to the drawn number of signal events 

\subsection{Systematic Uncertainties on the Neutrino Event Rate}
\begin{figure}[t]
\centering
\includegraphics[angle=0,width=8cm]{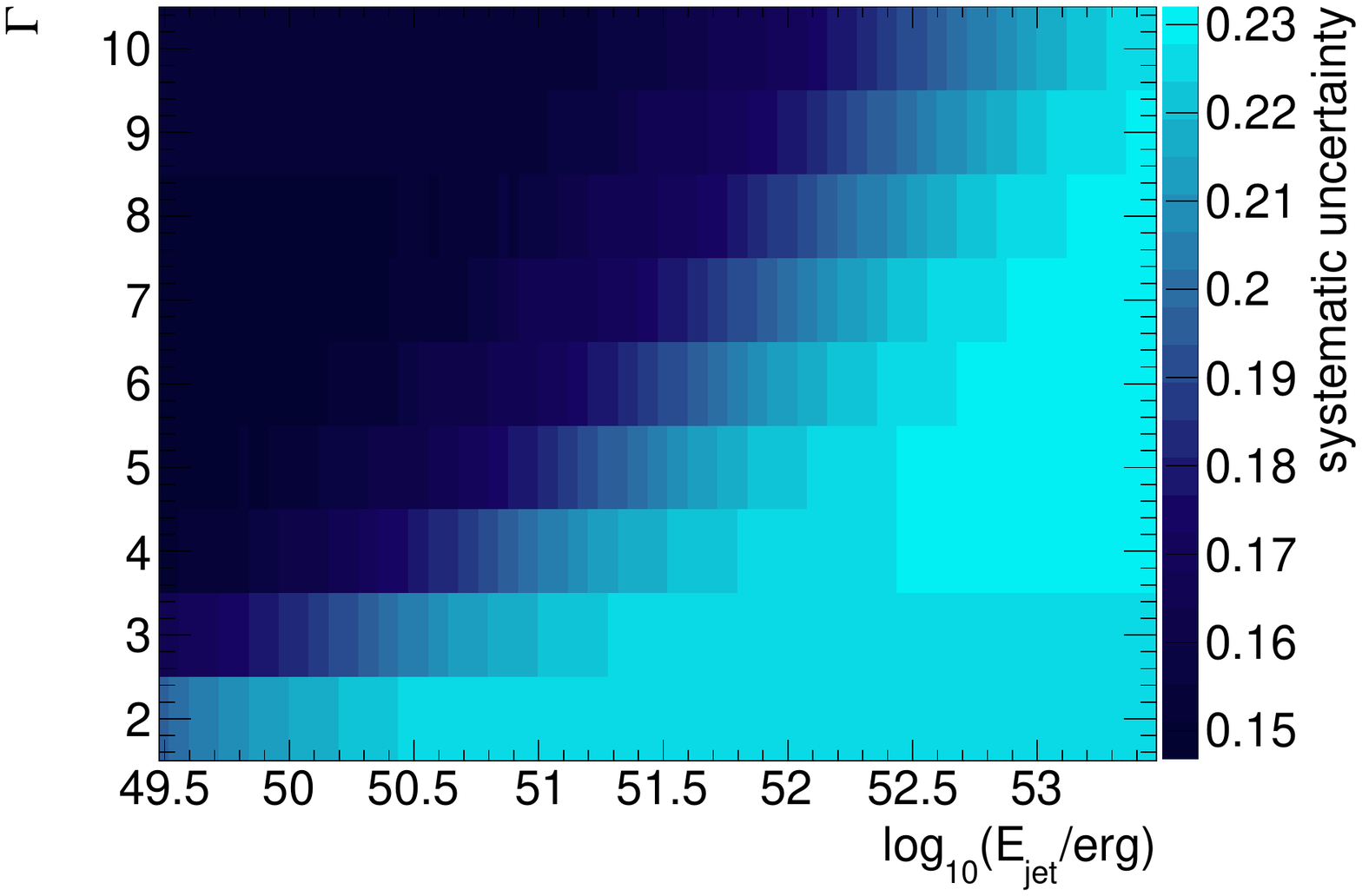}
\caption{Systematic uncertainties (relative to the predicted SN neutrino event rate) depending on model parameters $\Gamma$ and $E_{\rm{jet}}$.}
\label{pic:SysErrors}
\end{figure}

The main sources of systematic uncertainties arise in the description of the DOM efficiency and the photon propagation in ice. The systematic uncertainties due to photon propagation are evaluated by performing dedicated simulations with scattering and absorption coefficients varied within their uncertainties of \til$10\%$~\citep{IcePaper}. The maximum difference was found between the case where both scattering and absorption were increased by $10\%$ and the case where both were decreased by $10\%$. 
%Dedicated simulations varying the absorption and scattering coefficient as well as the DOM efficiency within their systematic uncertainties ($\sim 10\%$) were used to study the influence on the predicted event rate. It was found that the largest effect on the event rate (up to $13$\%) occurred when decreasing or increasing absorption and scattering coefficient at the same time. 
Varying the DOM efficiency resulted in a variation of the event rate of up to $18\%$. %Comparison between two different ice models showed variations in the event rate of up to $4\%$.
The neutrino cross section used in the neutrino simulation is based on CTEQ5 measurements, which are not up-to-date anymore. The latest cross section calculation by \cite{CooperSarkar} differ from the cross sections used in the IceCube neutrino simulation by up to $10$\% in the relevant energy regime. 
To first order, the neutrino rate depends linearly on the cross section. Folding the expected SN neutrino spectrum with an energy dependent correction factor allows us to calculate the effect on the neutrino event rate, which is up to $6$\%.\\
Finally, the uncertainty in the muon energy loss amounts to $1$\%, resulting in a $1$\% influence on the event rate~\citep{AMANDA5yearPS}.
The systematic uncertainties are energy dependent and the SN neutrino spectrum varies with the model parameters. Therefore, we calculate the quadratic sum of all listed systematic uncertainties for each combination of model parameters (see Fig.~\ref{pic:SysErrors}). 
Table~\ref{tab:SystematicErrorsIC} summarizes the systematic uncertainties considered in this analysis and their influence on the event rate for those model parameters, where the effect is largest.
\begin{table}[h!fbt]
 \caption[]{\sf{Systematic Uncertainties and their Influence on the Event Rates.}}
  \label{tab:SystematicErrorsIC}
 \begin{center}
  \begin{tabular}{ l c c }
    \toprule
    \textbf{Source of}                   & \textbf{Uncertainty} & \textbf{Average influence} \\
    \textbf{uncertainty}                 &                      & \textbf{on event rate} \\
%    \midrule
%    Ice model                                        &                       & $-1\%$, $+4\%$        \\
    \midrule
    DOM Efficiency                                   & $\pm10$\%             & $\pm18\%$        \\
    \midrule
    Photon propagation                               & $\pm10$\%             & $\pm13$\%        \\
    \midrule
    Neutrino cross section                           & $[-3\%,-10\%]$             & $-6$\%                     \\
    \midrule
    Muon energy loss                                 & $\pm1$\%              & $\pm1$\%                     \\
    \bottomrule
  \end{tabular}
 \end{center}
\end{table}
%\end{linenumbers}

\subsection{Systematic Uncertainties on the SN Sensitivity}
%\begin{linenumbers}
The number of expected observed SNe depends on the sensitivity of the telescopes. The estimate described in section~\ref{sec:OpticalCounterpart} yields the efficiency as a function of the SN peak magnitude. The photometric zero-points as determined from USNO A2.0 R-band magnitudes have typical systematic uncertainties of up to 0.30\,mag~(\cite{ROTSEOrphanAfterglow} and references therein). Shifting the efficiency curve by $\pm0.3$\,mag results in a variation of the expected number of SNe measured by ROTSE of $[-17.6\%$, $+26.6\%]$.\\
The expected number of accidentally found SNe depends on the overall core-collapse SN rate, which is assumed to be 1 SN per year within a sphere of radius $10$\,Mpc (continuum limit from~\cite{NearbySN}). The true SN rate might be higher since nearby SNe surveys tend not to target small galaxies. During the last decade $17$ SNe within $10$\,Mpc were observed \citep{FiveMegaton}. 
%Furthermore the local universe is not homogeneous and we are surrounded by an over density~\cite{OverDensity}. 
We assume a systematic uncertainty of $30\%$ due to inhomogeneity of the local universe and $30\%$ on the CCSN rate. Systematic uncertainty introduced by the scanning process are considered negligible.
%The visual scanning of the final candidates was carried out by three individual persons, who obtained similar results. All simulated SNe, which passed the computer selection, were identified in the scanning, i.e. the efficiency was $100\%$. Also the rate of false positives is expected to be small, because potential SN candidates would be inspected in detail (e.g. their light curve and the host galaxy) and measuring a spectrum at later times would be possible to ensure an unambiguous identification of the SN. We therefore assume that the systematic uncertainty introduced by the scanning process is small and neglect it.
\begin{table}[t!]%[h!fbt]
 \caption[]{\sf{Systematic Uncertainties}}
  \label{tab:SystematicErrorsROTSE}
 \begin{center}
  \begin{tabular}{c c c}
    \toprule
    \textbf{Source of}                   & \textbf{Uncertainty} & \textbf{Influence} \\
    \textbf{uncertainty}                   &                      & \textbf{on SN rate} \\

    \midrule
    Magnitude                                        & \multirow{2}{*}{$0.3$mag}              & \multirow{2}{*}{$-17.6\%$, $+26.6\%$}        \\
    measurement					     &                                        &					\\
    \midrule
    CC SN rate                                       & $\pm30\%$                & $\pm30$\%        \\
    \midrule
    Inhomogeneity of                                     & \multirow{2}{*}{$\pm30\%$}                & \multirow{2}{*}{$\pm30$\%}        \\
    the local universe                                   &                       &        \\
    \midrule
    \midrule
    Quadratic sum                                    &                          & $-45.9$, $+50.1$\%        \\
    \bottomrule
  \end{tabular}
 \end{center}
\end{table}
%\end{linenumbers}

\section{Results}
\label{sec:Results}
\begin{figure}[t]
\centering
\includegraphics[angle=0,width=10cm]{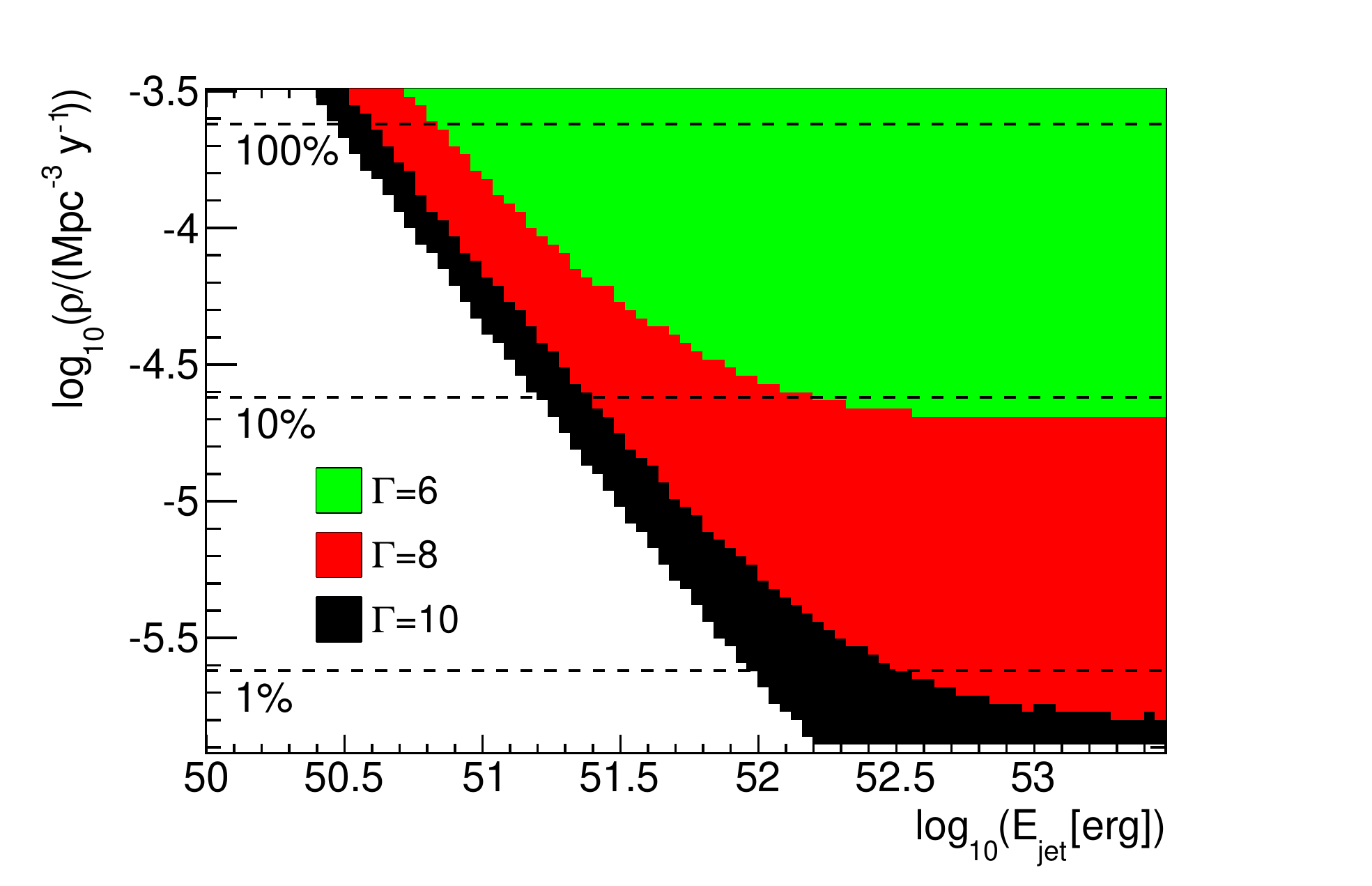}
\caption{Limits on the choked jet SN model~\cite{AndoBeacom} for different Lorentz boost factors $\Gamma$ as a function of the rate of SNe with jets $\rho$ and the jet energy $E_{\rm{jet}}$. The colored regions are excluded at 90\% confidence level.
%Blue: $\Gamma=10$. Red: $\Gamma=8$. Green: $\Gamma=6$. 
Horizontal dashed lines
indicate a fraction of SNe with jets of 100\%, 10\% or 1\% (relative to an assumed CCSN rate of 1 per year within a sphere of radius 10\,Mpc).}
\label{pic:limit}
\end{figure}

%\begin{linenumbers}
This paper presents the results from the analysis of data taking in the period of Dec. 16, 2008 to Dec. 31, 2009. IceCube was running initially in the IC40 configuration (Dec. 16, 2008 to May 20, 2009) and later in the IC59 configuration (May 20, 2009 to Dec. 31, 2009)\footnote{Note that the IceCube detector was running in the 40-string configuration already before Dec. 2008 and took data with the 59-string configuration also after Dec. 2009. The 2010 dataset is currently being analyzed.}.
Table~\ref{tab:multiplets} shows the number of detected and expected doublets and triplets for the IC40 and the IC59 datasets as well as the number of detected and expected optical SN counterparts.
The IceCube expectation based on a background-only hypothesis
was obtained from scrambled datasets.
 % In our signal estimation we have not accounted for mixed multiplets due to a single SN neutrino in coincidence with a background neutrino. While in principle, these can be indentified through an optical counterpart, we estimate the rate to be at most a few percent of the pure, signal-only multiplets. We hence negelted the extra contribution
To correctly incorporate detector asymmetries, seasonal variations 
and up-time gaps, we used the entire IC40 and IC59 datasets and exchanged the event directions randomly while 
keeping the event times fixed. For each scrambled dataset we obtain the number of doublets by 
comparing event directions. The number of doublets in both data sets (IC40 and IC59) shows a small excess, which corresponds to a $2.1$\,$\sigma$ effect and is thus not statistically significant. 
%To account for asymmetries in the azimuth distribution, we obtained 
%the doublets by comparing the detector coordinates of events in contrast to the online analysis where equatorial coordinates are used. 
%To locate an astrophysical source one has to use equatorial coordinates. But since we look for multiplets on
%a short time scale of 100~s, the rotation of the Earth during that time is negligible.\\
%\end{linenumbers}
\begin{table}[t!]%[h!fbt]
 \caption[]{\sf{Measured and Expected Number of Multiplets}}
  \label{tab:multiplets}
 \begin{center}
  \begin{tabular}{ l c c c c c}
    \toprule
                  & SN      & \multicolumn{2}{c}{Doublets}     & \multicolumn{2}{c}{Triplets}   \\
                  &         & IC40 & IC59                      & IC40 & IC59 \\
    \midrule
    measured      & 0       & 15   & 19                        &   0 & 0 \\
    \midrule
    expected      &  0.074  & 8.55 & 15.66                     &  0.0028 & 0.0040 \\
    \bottomrule
  \end{tabular}
 \end{center}
\end{table}
%\begin{linenumbers}
To estimate the expected number of randomly coincident SN detections, we assume a core-collapse SN rate 
of 1 per year within a sphere of radius 10\,Mpc, i.e.\
$2.4\times10^{-4}$\,y$^{-1}$\,Mpc$^{-3}$, and a Gaussian absolute magnitude distribution
with mean of $-18$\,mag and standard deviation of $1$\,mag~\citep{SNMag}. Based on the efficiency
estimated in section~\ref{sec:OpticalCounterpart} we can calculate the rate of core-collapse SNe that could be detected
by ROTSE, if it would continuously survey the full sky, $R_{\rm{CCSN}}^{\rm{ROTSE}} = 3823$\,y$^{-1}$. For this we integrated the CCSN rate over the accessible volume weighted with the efficiency displayed in Fig.~\ref{pic:ROTSEEffDist}.  The number of expected accidental SN detections (i.e. a SN detection in coincidence with a background neutrino multiplet) is
%\end{linenumbers}
\begin{equation}
\label{eq:BGSN}
 N_{\rm{SN,exp}} 
 %\mu^{SN}
 = \Delta T_{\rm{SN}} \cdot N_{\rm{alerts}} \cdot \frac{\Omega_{\rm{ROTSE}}}{\Omega_{\rm{sky}}} \cdot R_{\rm{CCSN}}^{\rm{ROTSE}}
                     = 0.074
\end{equation}
%\begin{linenumbers}
where $N_{\rm{alerts}}=17$ is the number of multiplet alerts followed-up by ROTSE, $\Omega_{\rm{ROTSE}} = 1.85^{\circ} \times 1.85^{\circ}$ is the solid angle covered by ROTSE's field of view and $\Omega_{\rm{sky}}=41253(^{\circ})^2$ is the all sky solid angle.
$\Delta T_{\rm{SN}}$ is the time window in which we accept a coincidence of neutrino and optical signals. It has to be larger than the uncertainty of the SN explosion time. 
%uncertainty of the SN explosion time, which is crucial
%to establish the coincidence of the optical and neutrino detection. 
In \cite{SNLightCurve} it is shown, that
the explosion time can be estimated with an accuracy of \til$1$\,day if early data are available. We choose to be conservative
and take $\Delta T_{\rm{SN}} = 5$\,days.
In total, 31 alerts were forwarded to the ROTSE telescopes. Five could not be observed because they
were too close to the Sun. For two alerts no good data could be collected. Seven alerts were discarded
because the corresponding fields were too close to the Galactic plane and hence too crowded. Thus 17 good optical datasets remained for
the analysis. The data were processed as described above. 
No optical SN counterpart was found in the data.
We calculate the limit on the AB05 model parameters following the description in appendix~\ref{appendixLimitCalculation} for the jet Lorentz boost factors $\Gamma = 6,8,10$ and in each case vary the jet energy $E_{\rm{jet}}$ and the rate of SNe with jets $\rho$. The algorithm was formulated prior to the start of the program.
The systematic uncertainties discussed in section~\ref{sec:SysErrors} are included in the limit calculation.
For each $\Gamma$-value the $90$\% confidence region in the $E_{\rm{jet}}$-$\rho$-plane is displayed in Fig.~\ref{pic:limit}. 
The colored regions are excluded with 90\% confidence. The limits %displayed in Fig.~\ref{pic:limit} 
include the optical information, i.e. that no optical counterpart was found. This improved the limit and allows tests of 5-25\% smaller CCSN rates. The largest improvement is obtained for small jet energies and large CCSN rates.
% \begin{figure}[h]
% \centering
% \includegraphics[angle=0,width=10cm]{images/limitFinalNew.pdf}
% \caption{Limits on the choked jet SN model for different boost Lorentz factors $\Gamma$ as a function of the rate of SNe with jets $\rho$ and the jet energy $E_{\rm{jet}}$. The colored regions are excluded at 90\% confidence level. 
% %Blue: $\Gamma=10$. Red: $\Gamma=8$. Green: $\Gamma=6$. 
% Horizontal dashed lines
% indicate a fraction of SNe with jets of 100\%, 10\% or 1\% (relative to an assumed CCSN rate of 1 per year within a sphere of radius 10\,Mpc).}
% \label{pic:limit}
% \end{figure}
The most stringent limit can be set for high Lorentz factors, while for small $\Gamma$ the constraints are weak. At $90\%$ confidence level, a
sub-population of SNe with typical values of $\Gamma$ and $E_{\rm{jet}}$ of $10$
and $3\times10^{51}$\,erg, respectively, does not exceed $4.2\%$.
%Less than $7.8\%$ of all SNe have a jet with $\Gamma=10$ and a typical GRB jet energy of $E_{\rm{jet}}=3\times10^{51}$\,erg. 
This is the first limit on CCSN jets using neutrino information.
%\end{linenumbers}

\section{Summary and Outlook}
\label{sec:Summary}
%\begin{linenumbers}
This first analysis, using the four ROTSE telescopes, proves the feasibility of
the program for follow-up observations triggered by neutrino multiplets detected by IceCube. 
%The optical follow-up program of IceCube neutrino multiplets realized with the four ROTSE telescopes
%proves the feasibility of the program. 
The technical challenge of analyzing neutrino data in real
time at the remote location of the South Pole and triggering optical telescopes has been solved.
First meaningful limits to the SN slow-jet hypothesis could be derived already after the first year of 
operation. Especially in cases of high Lorentz boost factors of $\Gamma=10$ stringent limits on the soft jet SN model are obtained.
\cite{SoderbergRadio} obtain an estimate on the fraction of SNe harboring a central engine, which powers a relativistic outflow, from a radio survey of type Ibc SNe. They conclude that the rate is about $1\%$, consistent with the inferred rate of nearby GRBs. Our approach is completely independent and for the first time directly tests hadronic acceleration in CCSNe, while the radio counterpart is sensitive to leptonic acceleration.\\
%\section{Outlook}
%\label{sec:Outlook}
The instrumented volume of IceCube has now 
%Meanwhile IceCube's volume has 
increased to a cubic kilometer yielding an increased sensitivity to 
high-energy neutrinos. In addition the live time is growing continuously. The delay of processing neutrino data at the South Pole has been reduced significantly from several 
hours to a few minutes. This results in the possibility of a very fast follow-up and allows the
detection of GRB afterglows, which fade rapidly below the telescope's detection threshold.\\
In addition, a single high-energy neutrino event trigger (in addition to the mutliplet trigger) is under development, which will further increase the sensitivity of the program especially for hard GRB neutrino spectra.\\
Because of the successful operation of the optical follow-up program with ROTSE, the program was extended
in August 2010 to the Palomar Transient Factory (PTF)~\citep{PTF1,PTF2}, which will provide deeper images and a fast
processing pipeline including a spectroscopic follow-up of interesting SN candidates. Furthermore, an X-ray follow-up by the Swift satellite~\citep{Swift} of the most significant multiplets %performed by the Swift satellite~\citep{Swift}
has been set up and started operations in February 2011.
%\end{linenumbers}

\begin{acknowledgements}
 
We acknowledge the support from the following agencies: U.S. National Science Foundation-Office of Polar Programs, U.S. National Science Foundation-Physics Division, University of Wisconsin Alumni Research Foundation, the Grid Laboratory Of Wisconsin (GLOW) grid infrastructure at the University of Wisconsin - Madison, the Open Science Grid (OSG) grid infrastructure; U.S. Department of Energy, and National Energy Research Scientific Computing Center, the Louisiana Optical Network Initiative (LONI) grid computing resources; National Science and Engineering Research Council of Canada; Swedish Research Council, Swedish Polar Research Secretariat, Swedish National Infrastructure for Computing (SNIC), and Knut and Alice Wallenberg Foundation, Sweden; German Ministry for Education and Research (BMBF), Deutsche Forschungsgemeinschaft (DFG), Research Department of Plasmas with Complex Interactions (Bochum), Germany; Fund for Scientific Research (FNRS-FWO), FWO Odysseus programme, Flanders Institute to encourage scientific and technological research in industry (IWT), Belgian Federal Science Policy Office (Belspo); University of Oxford, United Kingdom; Marsden Fund, New Zealand; Japan Society for Promotion of Science (JSPS); the Swiss National Science Foundation (SNSF), Switzerland; A. Gro\ss{} acknowledges support by the EU Marie Curie OIF Program; J. P. Rodrigues acknowledges support by the Capes Foundation, Ministry of Education of Brazil.\\
The ROTSE project is supported by NSF grant PHY-0801007 and NASA grant NNX08AV63G. We are grateful to Andre Phillips at Siding Spring Observatory, David Doss at the McDonald Observatory, Toni Hanke at the HESS Observatory and Tuncay \"{O}zi\c{s}ik at TUBITAK National Observatory for their invaluable efforts in maintaining the ROTSE telescopes.

\end{acknowledgements}

\appendix

\section{Limit Calculation}
\label{appendixLimitCalculation}
%\begin{linenumbers}
Motivated by the GRB-SNe connection, the soft SN jet model predicts the production of mildly
relativistic baryon loaded jets in core-collapse SNe. The resulting neutrino flux depends on the jet energy $E_{\rm{jet}}$ and the jet boost factor $\Gamma$. 
The rate of detectable SNe depends on the rate of core-collapse SNe producing a jet $\rho$.
In order to test the model we define a test statistic $\lambda$ consisting of an IceCube term $\lambda_{\rm{IC}}$ and a ROTSE term $\lambda_{\rm{ROTSE}}$.\\
The probability to detect $N_{k}$ or more multiplet events in IceCube with multiplicity $k$ over a background expectation of $\mu_k$ is 
given by the sum over Poisson probabilities:
%\end{linenumbers}
\begin{equation}
 P(N_k,\mu_k) = \sum_{i=N_k}^{\infty} \frac{\mu_k^i}{i!}e^{-\mu_k}.
\end{equation}
%\begin{linenumbers}
Combining all multiplicities and the two datasets yields the test statistic $\lambda_{\rm{IC}}$
%\end{linenumbers}
\begin{equation}
  \lambda_{\rm{IC}} = \prod_{k=2}^{\infty} P(N^{IC40}_k,\mu^{IC40}_k) \cdot P(N^{IC59}_k,\mu^{IC59}_k).
\label{equ:probBG}
\end{equation}
%\begin{linenumbers}
In addition to the IceCube information (i.e.\ number of doublets and multiplets of higher order) we add information obtained by the optical observations.
The probability to observe $N_{\rm{SN}}$ or more optical SN counterparts based on the expected number $\mu^{\rm{SN}}$ of accidentally observed SN in coincidence with an IceCube multiplet (given by Eq.~\ref{eq:BGSN})
is given by the sum of Poisson probabilities:
%\end{linenumbers}
\begin{equation}
 P_{\rm{SN}}(N^{\rm{SN}},\mu^{\rm{SN}}) = \sum_{i=N^{\rm{SN}}}^{\infty} \frac{\left(\mu^{\rm{SN}}\right)^{i}}{i!} e^{-\mu^{\rm{SN}}}.
\end{equation}
%\begin{linenumbers}
If one or more optical counterparts were observed the significance could be improved by adding neutrino timing information
as well as the distance information of the object found.
The probability $P_t$ to find a time difference of $\Delta t$ or smaller between the first and the last neutrino event in the multiplet (defined by the maximal temporal difference of $100$\,s between the neutrino arrival times) due to a background fluctuation assuming a uniform background is given by 
%\end{linenumbers}
\begin{equation}
\label{equ:pt}
 P_t = \frac{\Delta t}{100\textrm{\,s}}.
\end{equation}
%\begin{linenumbers}
Hence, assuming a generic SN prediction of a $10$\,s wide neutrino pulse results in a factor of 10 lower chance probability.\\
Taking into account the SN distance allows us to compute the probability $P_d$ to observe a background SN at a distance $d<d_{SN}$
%\end{linenumbers}
\begin{equation}
\label{equ:pd}
 %P_d = \frac{N_{\rm{SN,ROTSE}}(d<d_{\rm{SN}})}{N_{\rm{SN,ROTSE}}},
 P_d = \frac{R_{\rm{CCSN}}^{\rm{ROTSE}}(d<d_{\rm{SN}})}{R_{\rm{CCSN}}^{\rm{ROTSE}}},
\end{equation}
%\begin{linenumbers}
%where $N_{\rm{SN,ROTSE}}(d<d_{\rm{SN}})$ is the number of SN observable by the ROTSE telescopes within a sphere of radius $d_{\rm{SN}}$. $N_{\rm{SN,ROTSE}}$ is the total number of SN observable by ROTSE.
where $R_{\rm{CCSN}}^{\rm{ROTSE}}(d<d_{\rm{SN}})$ is the rate of SN observable by the ROTSE telescopes within a sphere of radius $d_{\rm{SN}}$. $R_{\rm{CCSN}}^{\rm{ROTSE}}$ is the total number of SN observable by ROTSE.
Accidental coincidences will be distributed following the SN rate (i.e. the square of the distance) folded with ROTSE's sensitivity as a function of distance $\varepsilon_{\rm{ROTSE}}(d)$ (see section~\ref{sec:OpticalCounterpart}). Signal events have a strong preference to close-by SNe, since only these will lead to a neutrino flux large enough to produce a detectable multiplet in IceCube. 
While ROTSE can only detect close-by SNe, more powerful telescopes can access a much larger volume and would essentially always detect a SN in their field of view. Hence, the additional factor $P_d$ becomes important to account for the SN distance.
The additional terms $P_t$ and $P_d$ for each observed SN light curve are combined with $P_{\rm{SN}}$ yielding the test statistic $\lambda_{\rm{ROTSE}}$.
%\end{linenumbers}
\begin{equation}
\lambda_{\rm{ROTSE}} = P_{\rm{SN}}(N^{\rm{SN}},\mu^{\rm{SN}}) \prod_{i=1}^{N_{\rm{SN}}} P_{t,i} \cdot P_{d,i}.
\end{equation}
%\begin{linenumbers}
Combining all information into one test statistic $\lambda$ yields:
%\end{linenumbers}
\begin{equation}
 \lambda = \lambda_{\rm{IC}} \cdot \lambda_{\rm{ROTSE}}.
\label{probModel}
\end{equation}
%\begin{linenumbers}
To obtain a proper confidence region for exclusion of the model we perform 10000 Monte Carlo (MC)
experiments for each combination of model parameters. 
%The mean number of expected SN neutrino multiplets $N^{\rm{IC40/IC59}}_{s,k}$ is given by the choice of model parameters. The average number of background multiplets $N^{\rm{IC40/IC59}}_{b,k}$ is known from scrambling. 
The model prediction, $N_{s,k}^{\rm{IC40/IC59}}$ and $N_{s}^{\rm{SN}}$, depends on the model parameters, $E_{\rm{jet}}$, $\Gamma$ and $\rho$. 
$N_{s,k}^{\rm{IC40/IC59}}$ is obtained from the neutrino signal simulation weighted with the corresponding AB05 spectrum. The neutrino spectrum varies with $E_{\rm{jet}}$ and $\Gamma$ as presented in section~\ref{sec:neutrinoFlux}. The number of predicted SNe, $N_{s}^{\rm{SN}}$, depends on the number of neutrino multiplets, i.e. number of telescope pointings, folded with the sensitivity of the telescope. In our signal estimation we have not accounted for mixed multiplets due to a single SN neutrino in coincidence with a background neutrino. While in principle, these can be indentified through an optical counterpart, we estimate the rate to be at most a few percent of the pure, signal-only multiplets. We hence neglected the extra contribution. The average number of background multiplets $N^{\rm{IC40/IC59}}_{b,k}$ is known from scrambling.\\
For each MC experiment the number of signal and background multiplets are drawn following a Poisson distribution. In case of the signal, the systematic uncertainties are included by smearing the Poisson mean, i.e. the Poisson mean is multiplied by a factor following a Gaussian distribution with mean one and a width given by the systematic uncertainties summarized in section~\ref{sec:SysErrors}.\\
%In a similar way the number of signal and background SN counterparts is drawn in every MC experiment. 
If an optical counterpart is drawn in the MC simulation ($N_{s}^{\rm{SN}}\geq1$) we calculate the additional terms $P_t$ and $P_d$ following equations~\ref{equ:pt} and~\ref{equ:pd}. The time difference between the SN neutrinos is set to $\Delta t = 10$\,s and the SN distance is thrown following a spatially isotropic distribution folded with ROTSE's efficiency. 
For each MC experiment $\lambda$ is calculated following equation~\ref{probModel}. The fraction of MC experiments
resulting in a smaller value of $\lambda$ than that of the data sample (i.e. fraction of outcomes of the MC experiment which show worse agreement with the background-only hypothesis than the actual measurement) is the desired confidence level for the exclusion of this combination of model parameters.
% The model prediction, $N_{s,k}^{\rm{IC40/IC59}}$ and $N_{s}^{\rm{SN}}$, depend on the model parameters, $E_{\rm{jet}}$, $\Gamma$ and $\rho$. 
% $N_{s,k}^{\rm{IC40/IC59}}$ is obtained from the neutrino signal simulation weighted with the corresponding AB05 spectrum. The neutrino spectrum varies with $E_{\rm{jet}}$ and $\Gamma$ as presented in section~\ref{sec:neutrinoFlux}. The number of predicted SNe, $N_{s}^{\rm{SN}}$, depends on the number of neutrino multiplets, i.e. number of telescope pointings, folded with the sensitivity of the telescope.
%\end{linenumbers}

%\input{appendixDataMC}

\bibliographystyle{aa} % style aa.bst
\bibliography{OFUPaper}

%\clearpage

%\pagebreak

%\clearpage

\end{document}